\pdfoutput=1
\documentclass{article}
\usepackage{arxiv}
\usepackage[utf8]{inputenc} 
\usepackage[T1]{fontenc}    
\usepackage{hyperref}       
\usepackage{url}            
\usepackage{booktabs}       
\usepackage{amsfonts}       
\usepackage{nicefrac}       
\usepackage{microtype}      
\usepackage{lipsum}
\usepackage{graphicx}
\usepackage{amsmath}
\usepackage{esvect}
\usepackage{verbatim}
\usepackage{xcolor}
\usepackage[misc]{ifsym}
\usepackage{etoolbox}
\usepackage{kantlipsum}
\usepackage[inline]{enumitem}
\usepackage{scrtime}
\usepackage[dont-mess-around]{fnpct}
\usepackage[numbers]{natbib}
\usepackage{float}

\newcommand*{\affmark}[1][*]{\textsuperscript{#1}}

\title{Review on Learning and Extracting Graph Features for Link Prediction}

\author{
Ece C. Mutlu\thanks{Represents equal contribution to this work.}\\
  Department of Industrial Engineering\\
  University of Central Florida\\
  \texttt{ece.mutlu@ucf.edu} \\
   \And
 Toktam A. Oghaz\affmark[*]\\
  Department of Computer Science\\
  University of Central Florida\\
  \texttt{toktam@cs.ucf.edu} \\
   \And
 Amirarsalan Rajabi\affmark[*]\\
  Department of Computer Science\\
  University of Central Florida\\
  \texttt{amirarsalan@knights.ucf.edu} \\
  \And
  Ivan Garibay\textsuperscript{\Letter}\\
  Department of Industrial Engineering\\
  Department of Computer Science\\
  University of Central Florida \\
  \texttt{igaribay@ucf.edu} \\
}

\begin{document}
\maketitle
\begin{abstract}
Link prediction in complex networks has attracted considerable attention from interdisciplinary research communities, due to its ubiquitous applications in biological networks, social networks, transportation networks, telecommunication networks, and, recently, knowledge graphs. Numerous studies utilized link prediction approaches in order sto find missing links or predict the likelihood of future links as well as employed for reconstruction networks, recommender systems, privacy control, etc. This work presents an extensive review of state-of-art methods and algorithms proposed on this subject and categorizes them into four main categories: similarity-based methods, probabilistic methods, relational models, and learning-based methods. Additionally, a collection of network data sets has been presented in this paper, which can be used in order to study link prediction. We conclude this study with a discussion of recent developments and future research directions.
\end{abstract}


\section{Introduction}
\label{sec:Intro}

Online social networks \cite{ahuja2019using}, biological networks, such as protein--protein interactions and genetic interactions between organisms \cite{sulaimany2018link}, ecological systems of species, knowledge graphs \cite{wang2017knowledge}, citation networks \cite{zhou2018h}, and social relationships of users in personalized recommender systems \cite{ebrahimi2016personalized}, are all instances of graphs of complex interactions, which are also referred to as complex networks. While these networks are almost always dynamic in nature, a vital query is how they change over time. More specifically, what are the future associations between entities in a graph under investigation. The problem of link prediction in graphs is one of the most interesting and long-standing challenges. Given a graph, which is an abstraction of relationships among entities of a network, link prediction is to anticipate future connections among entities in the graph, with respect to its current state. Link prediction models might 
\begin{enumerate*}[label=(\roman*),leftmargin=21pt,labelsep=7pt]
\item exploit the similarity metrics as the input features, 
\item embed the nodes into a low dimensional vector space while preserving the topological structure of the graph, or 
\item combine the information that is derived from the two aforementioned points, with the node attributes available from the data set. 
\end{enumerate*}

All of these models rely on the hypothesis that higher similarity between nodes results in a higher probability of connection \cite{martinez2017survey}.

Applications of link prediction include analyzing user--user and user--content recommendations in online social networks \cite{ebrahimi2016personalized,song2009scalable,hristova2016multilayer,jalili2017link}, reconstruction of the PPI (protein--protein interaction) network and reducing the present noise \cite{lei2013novel,iakovidou2010multiway,airoldi2006mixed}, hyper-link prediction \cite{zhang2018beyond}, prediction of transportation networks \cite{ma2017intercity}, forecasting the behavior of terrorism campaigns and social bots \cite{desmarais2013forecasting,heidaribots}, reasoning and sensemaking in knowledge graphs \cite{xiao2016one}, and knowledge graph completion while using data augmentation with Bidirectional Encoder Representations from Transformers (BERT) \cite{devlin2018bert,zhao2020learning}. Link prediction in these applications has been mostly investigated through unsupervised graph representation and feature learning methods that are based on the node (local) or path (global) similarity metrics that evaluate the neighboring nodes. Common neighbors, preferential attachment, Jaccard, Katz, and Adamic Adar are some of the most widely used similarity metrics that measure the likelihoods of edge associations in graphs. While these methods may seem to be dated, they are far from being obsolete. Despite the fact that these methods do not discover the graph attributes, they have remained popular for years, due to their simplicity, interpretability, and scalability. Probabilistic models, on the other hand, aim to predict the likelihood of future connections between entities in an evolving dynamical graph with respect to the current state of the graph. Another context under which the problem of link prediction is raised is relational data \cite{heckerman1995learning,friedman1999learning,getoor2007probabilistic,getoor2002learning}. In this context, when considering the relational data set in which objects are related to each other, the task of link prediction is to predict the existence and type of links between pairs of objects \cite{al2011survey}. However, the availability of labeled data allows for the supervised machine learning algorithms to provide new solutions for the link prediction task, including neural network-based methods for link prediction \cite{zhang2018link}, which allow for learning a suitable heuristic than assuming strong relationships among vertices. 

Similar surveys on the topic of link prediction exist, and this survey has benefited from them. The work of \cite{kumar2020link} provides a comprehensive review of the problem of link prediction within different types of graphs and the applications of different algorithms. Other related review papers on this topic include the works of  \cite{nickel2015review} and \cite{lu2011link}. The work of \cite{lu2011link} reviews the progress of link prediction algorithms from a physical perspective, applications, and challenges for this line of research. While some of these reviews only focus on a specific set of methodologies that are proposed for link prediction, such as the work of \cite{nickel2015review}, which presents an extensive review on relational machine learning algorithms, specifically designed for knowledge graphs, some important related methodologies are overlooked in the aforementioned studies. For instance, \cite{lu2011link} does not discuss some important graph feature learning and neural network-based techniques that have been recently developed. Our effort has been to provide a review that includes the most recent approaches for the problem of link prediction that demonstrate promising results, but are not fully covered by exceptional similar surveys, such as the works \mbox{of \cite{lu2011link,kumar2020link,nickel2015review}}.
Thus, we believe that our study provides comprehensive information on the topic of link prediction for large networks, and it can help to discover the most related link prediction algorithms that are deliberately categorized into the proposed taxonomy. This study reviews similarity-based methods, including local, global, and quasi-local approaches, probabilistic and relational methods as unsupervised solutions to the link prediction problem, and, finally, learning-based methods, including matrix factorization, path and walk based link prediction models, and using neural networks for link~prediction.

\section{Background}
\label{sec: problemdescription}

\begin{figure}[H]
    \centering
    \includegraphics[width=300pt]{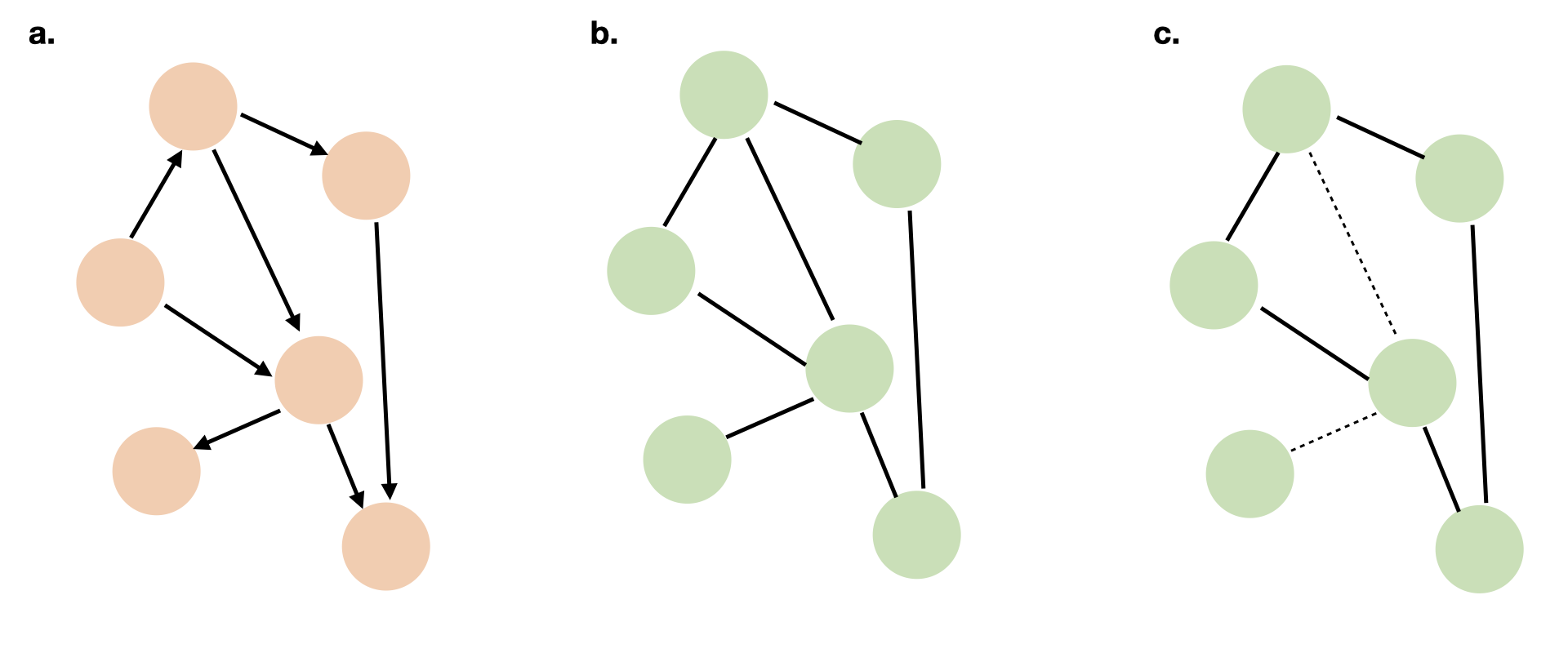}
    \caption{Imaginary representation of (\textbf{a}). Directed whole graph (\textbf{b}). Undirected whole graph (\textbf{c}). Undirected training graph.}
    \label{fig:my_label}
\end{figure}

A graph (complex network), denoted as  $G=\langle V, E \rangle$, can be defined as the set of vertices (nodes) $V$, and the interactions among pairs of nodes, called links (edges) $E$, at a particular time $t$. It should be noted that in this problem setting, self-connections, and multiple links between nodes are not allowed and, accordingly, are not taken into account in the majority of link prediction problem settings~\cite{lu2011link}. The main idea behind applying feature extraction or feature learning-based methods for the link prediction problem is to use the present information regarding the existing edges in order to predict the future or missing link that will emerge at time $t^{'}>t$. The types of graphs can be classified into two main categories according to the direction of the information flow between interacted nodes; directed and undirected graphs. Although many of the discussed methods in the next sections of this paper can provide solutions to the link prediction problem in directed graphs, the majority of the reviewed methods in this survey address the problem of link prediction for undirected graphs. The difference between the link prediction problem for these two graph categories arises from the additional information that is required for the directed graphs. This information refers to the origin of the associated link in directed graphs, in which $\langle v_x,v_y \rangle$ conveys the existence of a directed edge from node $v_x$ to $v_y$ and $\langle v_x,v_y \rangle \ne \langle v_y,v_x \rangle$ \cite{wasserman1994social}. However, edges in undirected graphs have no orientation, and the relations among the node pairs are reciprocal. The set of nodes that are connected to node $v_x \in V$ are known as the ``neighbors" of $v_x$, denoted as $\Gamma(v_x) \subseteq V$, and the number of edges that are connected to the node $v_x$ is referred to as $|\Gamma(v_x)|$. Link prediction algorithms necessitate training and test sets to be compared in the case of model performance, similar to other machine learning methods. However, one cannot know the future links of a graph at time $t^{'}$, given the current graph structure. Therefore, a fraction of links from the current graph structure is deleted (Figure \ref{fig:my_label}), and taken as the test set; whereas, the remaining fraction of edges in the graph is used for the training purpose. A reliable link prediction approach should provide higher probabilities for the edges that belong to the set of true positives than the set of nonexistent edges \cite{wang2018link}. Apparently, by treating the link prediction task as a binary classification problem, conventional evaluation metrics of binary classification in machine learning can be applied in order to evaluate the performance of link prediction. Within the context of the confusion matrix, TP (True Positive), FP (False Positive), TN (True Negative), and FN (False Negative) metrics can be used in order to assess performance.  In this context, sensitivity, specificity, precision, and accuracy are computed, as follows (\cite{manning2008introduction}):
\begin{equation}
\begin{split}
    \text{Sensitivity (Recall)} &= \frac{\text{TP}}{\text{TP} + \text{FN}}\\
    \text{Specificity} &= \frac{\text{TN}}{\text{TN}+\text{FP}}\\
    \text{Precision} &= \frac{\text{TP}}{\text{TP}+\text{FP}}\\
    \text{Accuracy} &= \frac{\text{TP}+\text{TN}}{\text{TP}+\text{TN}+\text{FP}+\text{FN}}\\
\end{split}
\end{equation}

The most common standard metric that is used to quantify the performance of the link prediction algorithms is "the area under the receiver operating characteristic curve (AUC)" \cite{hanley1982meaning}. The AUC value represents the probability that a randomly selected missing link between two nodes is given a higher similarity score than the randomly selected pair of unconnected links. The algorithmic calculation of AUC is given by:
\begin{equation}
    \text{AUC} = \frac{n'+0.5n''}{n}
\end{equation}
where $n$ is the number of total independent comparisons and $n'$ is the number of comparisons in which the missing link has a higher score than the unconnected link, while $n''$ is the number of comparisons when they show equal scores.

One of the applications of link prediction is in recommender systems \cite{chen2005link,ebrahimi2016personalized} that exploit information on users' social interactions in order to find their desired information according to their interests and preferences. Therefore, within this context, the following evaluation metrics are also used \cite{Gray2009,voorhees2000overview, Zhang2009}:
\begin{equation}
\begin{split}
    \text{Precision\: at} \: n &= \frac{r}{n}\\
    \text{Recall\: at} \: n &= \frac{r}{R}\\
\text{Mean\: Reciprocal \: Rank} &= \frac{1}{|Q|} \sum_{i=1}^{|Q|}\frac{1}{rank_i}\\
\text{Average\: Precision} &= \sum_{k=1}^{n}P(k)\Delta r(k)\\
\text{Mean\: Average\: Precision} &= \frac{\sum_{q=1}^{Q}AveP(q)}{Q}
\end{split}
\end{equation}

In the above equations, Precision at $n$ shows the number of relevant results ($r$) among the top $n$ results or recommendations.  In Recall at $n$, $R$ presents the total number of relevant results. In order to calculate Mean Reciprocal Rank, first, the inverse of the ranking of the first correct recommendation is calculated ($\frac{1}{rank_{i}}$) and, then, an average over the total queries ($Q$) is taken. In order to calculate Average Precision, Precision at a threshold of $k$ in the list is multiplied by the change in recall from items $k-1$ to $k$, and this process is summed over all of the positions in the ranked sequence of documents. The Mean Average Precision is then the average of all Average Precisions over total queries (\emph{Q}).

\begin{figure}[H]
    \centering
    \includegraphics[width=0.8\textwidth]{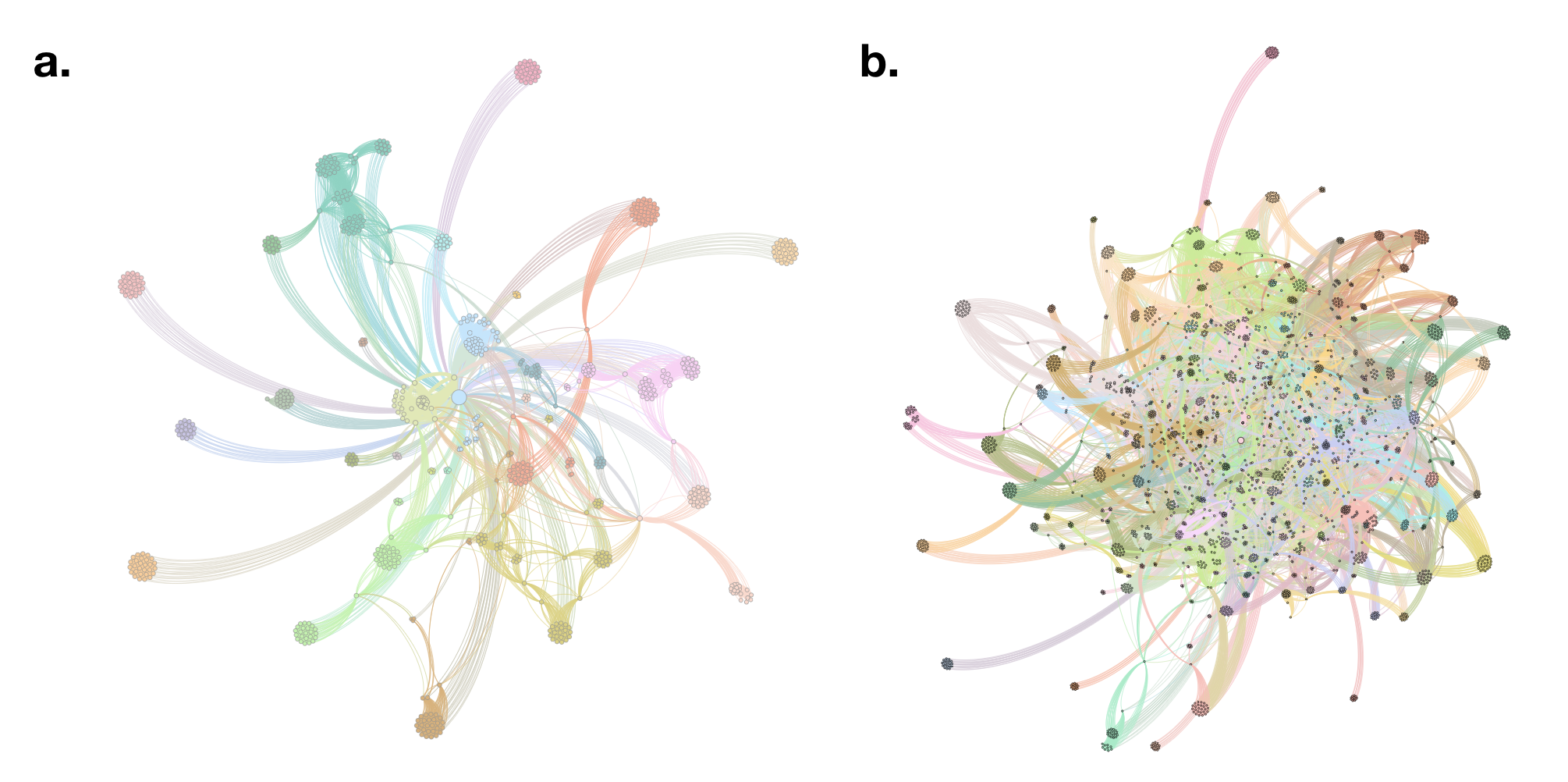}
    \caption{\textls[-5]{Hashtag co-occurrence graph (via Gephi) of (\textbf{a}). $\#$askmeanything (757 nodes and 16046 edges), (\textbf{b}). $\#$lovemylife (2748 nodes and 63413 edges). Each network is colored by the modularity ranking.}}
    \label{fig:gephi}
\end{figure}

In order to provide a few visualization examples for complex networks, Figure \ref{fig:gephi} demonstrates the network structure of the two different hashtag co-occurrence graphs ($\#$askmeanything and $\#$lovemylife) of the Instagram posts from 04/01/2020 to 04/08/2020. These two different figures clearly demonstrate the variability of the network structure, even in the same fields, i.e., Figure \ref{fig:gephi}a. shows different sub-communities with its more sparse structure, while Figure \ref{fig:gephi}b. represents a densely connected network example.

\section{Similarity Based Methods}
Similarity-based methods, which mainly focus on the topological structure of the graph, are the most straightforward and oldest link prediction metrics. These methods try to figure the missing links out by assigning similarity score, $s_{(v_x,v_y)}$, between node pairs ($v_x$ and $v_y$) using the structural property of the graphs. These methods can be investigated under three main categories: local, quasi-local, and global approaches.

\subsection{Local Similarity-Based Approaches}
\label{sec:Local}
Local similarity-based approaches are based on the assumption that, if node pairs have common neighbor structures, they will probably form a link in the future. Because they only use local topological information based on neighborhood-related structures rather than considering the whole network topology, they are faster than the global similarity-based approaches. Many studies also showed their superior performance, especially on the dynamic networks \cite{liben2007link}. However, they are restricted to compute the similarity of all possible combinations of the node pairs, since they only rank similarity between close nodes with a distance of less than two.

\subsubsection{Common Neighbors (CN)} CN is one of the most extensive information retrieval metrics for link prediction tasks, due to its high efficiency, despite its simplicity. The idea behind CN is very intuitive; the probability of being linked for two nodes in the future is affected by the number of their common neighboring nodes, i.e., two nodes will highly probably establish a link if they have more shared nodes. The score of this metric can be defined, as follows:
\begin{equation}
    s_{(v_x,v_y)}^{CN} = |\Gamma(v_x) \cap \Gamma(v_y)|
\end{equation}
where $\Gamma(.)$ represents the set of adjacent nodes. 

It should be noted that the resulting score using CN is not normalized, and only shows the relative similarity of different node-pairs by considering shared nodes between them. Newman used CN in order to show that the probability of collaboration between two scientists in the future can be estimated by their previous common collaborators \cite{newman2001clustering}.   

\subsubsection{Jaccard Index (JC)} 
The metric not only takes the number of common nodes into account as in CN, but it also normalizes it by considering the total set of numbers of shared and non-shared neighbors. The equation of this score that is proposed by Jaccard \cite{jaccard1901etude} is:
\begin{equation}
    s_{(v_x,v_y)}^{JC} = \frac{|\Gamma(v_x) \cap \Gamma(v_y)|}{|\Gamma(v_x) \cup \Gamma(v_y)|}
\end{equation}

\subsubsection{Salton Index (SL)} 
SL is the metric that is also known as cosine similarity. It calculates the cosine angle between the two columns of the adjacency matrix and it is identified as the ratio of the number of shared neighbors of $v_x$ and $v_y$ to the square root of inner-product of their degrees \cite{salton1986introduction}, as follows:  
\begin{equation}
    s_{(v_x,v_y)}^{SL} = \frac{|\Gamma(v_x) \cap \Gamma(v_y)|}{\sqrt{|\Gamma(v_x)|.|\Gamma(v_y)}}
\end{equation}

Wagner \& Leydesdorff \cite{wagner2005mapping} showed that SI is an efficient metric, especially when the aim is to visualize the constructional pattern of relations in a graph.

\subsubsection{S{\o}rensen Index (SI)} 
The index, which is very similar to JC, is generated to make a comparison between different ecological samples \cite{sorensen1948method}, such that: 
\begin{equation}
    s_{(v_x,v_y)}^{SI} = \frac{|\Gamma(v_x) \cap \Gamma(v_y)|}{|\Gamma(v_x)|+|\Gamma(v_y)|}
\end{equation}

The difference in using the summation of the degrees instead of the size of the union of their neighbors makes SI less outlier sensitive when compared to JC \cite{mccune2002analysis}.

\subsubsection{Preferential Attachment Index (PA)} 
Motivated by the study by Barabasi \& Albert \cite{barabasi1999emergence}, new nodes joining the network are more likely to connect with the nodes with higher connections (hub) than the nodes with lower degrees, PA can be formulated as:
\begin{equation}
s_{(v_x,v_y)}^{PA} = |\Gamma(v_x)|.|\Gamma(v_y)|
\end{equation}

\subsubsection{Adamic-Adar Index (AA)}
The metric is employed for the necessity of the comparison of two web-pages by Lada Adamic and Eytan Adar \cite{adamic2003friends}. It simply uses the idea of giving more weight to the relatively fewer common neighbors, such that:
\begin{equation}
s_{(v_x,v_y)}^{AA} = \sum_{v_z \in (\Gamma(v_x) \cap \Gamma(v_y))}^{} \frac{1}{\log|\Gamma(v_z)|}
\end{equation}
where $v_z$ refers to a common neighbor for nodes $v_x$ and $v_y$ (connected/linked to both). 

Although this metric has similarities to CN, the vital difference is that the logarithm term penalizes the shared neighbors of the two corresponding nodes. It should be noted that while the other metrics include only two nodes ($v_x$ and $v_y$) and/or their degrees in their equations so far, AA also relates familiar neighbors ($v_z$) to these two nodes ($v_x$ and $v_y$). 
\subsubsection{Resource Allocation Index (RA)} 
Motivated by the physical process of resource allocation, a very similar metric to AA was developed by Zhou et al. \cite{zhou2009predicting} which can be formulated as: 
\begin{equation}
s_{(v_x,v_y)}^{RA} = \sum_{v_z \in (\Gamma(v_x) \cap \Gamma(v_y))}^{} \frac{1}{|\Gamma(v_z)|}
\end{equation}

The difference in the denominator ($|\Gamma(v_z)|$) of RA rather than its logarithm (${log|\Gamma(v_z)|}$) as in AA penalizes the contribution of common neighbors more. Many studies show that this discrepancy is insignificant, and the resulting performances of these two metrics are very similar when the average degree of the network is low; however, RA is superior when the average degree is high \cite{wang2015link}. 

\subsubsection{Hub Promoted Index (HP)}
The index is proposed for assessing the similarity of the substrates in the metabolic networks \cite{ravasz2002hierarchical}, and it can be defined, as follows: 
\begin{equation}
s_{(v_x,v_y)}^{HP} = \frac{|\Gamma(v_x) \cap \Gamma(v_y)|}{\min(|\Gamma(v_x)|,|\Gamma(v_y)|)}  
\end{equation}

HP is determined by the ratio of the number of common neighbors of both $v_x$ and $v_y$ to the minimum of degrees of $v_x$ and $v_y$. Here, link formation between lower degree nodes and the hubs is more promoted, while the formation of the connection between hub nodes are demoted \cite{martinez2017survey}.

\subsubsection{Hub Depressed Index (HD)} 
The totally opposite analogy of HP is also considered by L\"{u} and Zhou \cite{lu2011link}, and it is determined by the ratio of the number of common neighbors of both $v_x$ and $v_y$ to the maximum of degrees of $v_x$ and $v_y$. Here, the link formation between lower degree nodes and link formation between hubs is promoted. However, the connection between hub nodes and lower degree nodes are demoted, such~that: 
\begin{equation}
s_{(v_x,v_y)}^{HD} = \frac{|\Gamma(v_x) \cap \Gamma(v_y)|}{\max(|\Gamma(v_x)|,|\Gamma(v_y)|)}    
\end{equation}

\subsubsection{Leicht-Holme-Newman Index (LHN)} 
The index, which is very similar to SI, is defined as the ratio of the number of shared neighbors of $v_x$ and $v_y$ to the product of their degrees (the expected value of the number of paths of length between them) \cite{leicht2006vertex}. It can be represented by:
\begin{equation}
s_{(v_x,v_y)}^{LHN} = \frac{|\Gamma(v_x) \cap \Gamma(v_y)|}{|\Gamma(v_x)|.|\Gamma(v_y)|} 
\end{equation}

The only difference in the denominator as compared to SI shows that SI always assigns a higher score than LHN, i.e., $|\Gamma(v_x)|.|\Gamma(v_y)| \leq |\Gamma(v_x) \cap \Gamma(v_y)|$ .

\subsubsection{Parameter Dependent Index (PD)} 
Zhou et al. \cite{zhu2012uncovering} proposed a new metric in order to improve the prediction accuracy for popular links and unpopular links. PD can be defined as: 
\begin{equation}
   s_{(v_x,v_y)}^{PD} = \frac{|\Gamma(v_x) \cap \Gamma(v_y)|}{{|\Gamma(v_x)|.|\Gamma(v_y)|}^\beta},
\end{equation}
where $\beta$ is a free parameter and it can be tuned to the topology of the graph. One can easily recognizes that PD is degraded to CN, SL, and LHN when $\beta=0$, $\beta=0.5$, and $\beta=1$, respectively.

\subsubsection{Local Affinity Structure Index (LAS)}
LAS shows the affinity relationship between a pair of nodes and their common neighbors. The hypothesis is that a higher affinity of two nodes and their common neighbors increases the probability of getting connected \cite{sun2017improved}, such as: 
\begin{equation}
s_{(v_x,v_y)}^{LAS} = \frac{|\Gamma(v_x) \cap \Gamma(v_y)|}{|\Gamma(v_x)|} + \frac{|\Gamma(v_x) \cap \Gamma(v_y)|}{|\Gamma(v_y)|}
\end{equation}

\subsubsection{CAR-Based Index (CAR)}
When a node interacts with another neighbor node, it is called a first-level neighborhood; whereas, the interaction between the first-level neighbor node and its neighbor node is called the second-level neighborhood for the seed node. According to the local community paradigm (LCP) of Cannistraci \cite{cannistraci2013link}, the researchers mostly consider the first-level neighborhood, because the second-level neighborhood is noisy; however, the second-level neighborhood carries essential information regarding the topology of the network. Therefore, CAR filters these noises and considers nodes that are interlinked with neighbors mostly. The similarity metric can be calculated, as follows:
\begin{equation}
    s_{(v_x,v_y)}^{CAR} = |\Gamma(v_x) \cap \Gamma(v_y)|\sum_{v_z \in (\Gamma(v_x) \cap \Gamma(v_y))} {\frac{|\Gamma(v_z)|}{2}}.
\end{equation}

\subsubsection{The Individual Attraction Index (IA)}
Dong et al. \cite{dong2011link} proposed an index that relates not only to the common neighbors of the nodes individually, but also the effect of the sub-network created by those. The IA score can be formulated as:
\begin{equation}
 s_{(v_x,v_y)}^{IA} = \sum_{v_z \in (\Gamma(v_x) \cap \Gamma(v_y))}^{} \frac{|\Gamma(v_x) \cap \Gamma(v_y) \cap \Gamma(v_z)|+2}{|\Gamma(v_z)|}. 
\end{equation}

Because IA considers the existence of links between all common neighbors, the algorithm is very time-consuming. Therefore, a simpler alternative is also proposed as:
\begin{equation}
     s_{(v_x,v_y)}^{IA*} = \sum_{v_z \in (\Gamma(v_x) \cap \Gamma(v_y))}^{} \frac{|\Gamma(v_x) \cap \Gamma(v_y)|+2}{|\Gamma(v_x) \cap \Gamma(v_y)|.|\Gamma(v_z)|}
\end{equation}

\subsubsection{The Mutual Information Index (MI)}
This method examines the link prediction problem while using information theory, and it measures the likelihood of conditional self-information when their common neighbors are known \cite{tan2014link}, and formulated as:
\begin{equation}
s_{(v_x,v_y)}^{MI} = -I(e_{v_x,v_y}|v_z),
\end{equation}
where $v_z \in \Gamma(v_x) \cap \Gamma(v_y)$ and $I(.)$ is the self-information function for a node and it can be calculated by (\ref{eq: selfinfo}). Here, $I(e_{v_x,v_y}|v_z)$ means conditional mutual  self-information of the existence of a link between $v_x$ and $v_y$ and their shared set of neighbors. The smaller value of $s_{(v_x,v_y)}^{MI}$ means the higher likelihood to be linked. If all of the link between common neighbors be independent of each other, the self-information of that node pair can be calculated as \cite{martinez2017survey}:
\begin{equation}
\label{eq: selfinfo}
    I(e_{v_x,v_y}|v_z)=\log2\frac{|\{e_{v_x,v_y}:v_x,v_y \in \Gamma(v_z), e_{v_x,v_y} \in E \}|}{\frac{1}{2}|\Gamma(v_z)|(|\Gamma(v_z)|-1)}.
\end{equation}

\subsubsection{Functional Similarity Weight (FSW)}

This index is first used by Chou et al. in order to understand the similarity of physical or biochemical characteristics of proteins \cite{chua2006exploiting}. Their motivation is based on the Czekanowski--Dice distance that is used in \cite{brun2003functional} in order to estimate the functional similarity of proteins. This score can be defined as:
\begin{equation}
   s_{(v_x,v_y)}^{FSW} = \bigg(\frac{2|\Gamma(v_x) \cap \Gamma(v_y)|}{|\Gamma(v_x) - \Gamma(v_y)| + 2|\Gamma(v_x) \cap \Gamma(v_y)| + \beta}\bigg)^2 .
\end{equation}

Here, $\beta$ is used to penalize the nodes with very few common neighbors, and it is defined as:
\begin{equation}
  \beta = \max(0, |\Gamma_{avg}|-(|\Gamma(v_x) - \Gamma(v_y)|)+ (|\Gamma(v_x) \cap \Gamma(v_y)|)), 
\end{equation}
where $|\Gamma_{avg}|$ is the average number of neighbours in the network.

\subsubsection{Local Neighbors Link Index (LNL)}

Motivated by the cohesion between common neighbors and predicted nodes, both attributes, and topological features are examined in \cite{yang2015new}, as:
\enlargethispage{0.5cm}
\begin{equation}
s_{(v_x,v_y)}^{LNL} = \sum_{v_z \in (\Gamma(v_x) \cap \Gamma(v_y))}{w(v_z)},
\end{equation}
where $w(v_z)$ is the weight function that can be measured by:
\vspace{12pt}
\begin{equation}
    w(v_z) = \frac{\sum_{v_u \in \Gamma(v_x) \cup v_x}{\delta(v_z,v_u)}+\sum_{v_v \in \Gamma(v_y) \cup v_y}{\delta(v_z,v_y)}}{|\Gamma(v_z)|}.
\end{equation}

Here, $\delta(a,b)$ is a boolean variable that is equal to 1 if there exists a link between a and b; otherwise, it equals to 0.

\subsection{Global Similarity-Based Approaches}
\label{sec:Global}
Global similarity-based approaches, contrary to local ones, use the whole topology of the network to rank the similarity between node pairs; therefore, they are not limited to measure the similarity between nodes that are locating far away from each other. Although considering the whole topology of the network gives more flexibility in link prediction analysis, it also increases the algorithm's time complexity. Because an ensemble of all paths between node pairs is used, they can also be called path-based methods.

\subsubsection{Katz Index (KI)} 
The metric, which is defined by Katz \cite{katz1953new}, sums over the sets of paths and is exponentially damped by length to be counted more intensively with shorter paths. This index can be formulated with a vector space:
\begin{equation}
  s_{(v_x,v_y)}^{KI} = \sum_{i=1}^{\infty}{\beta^i . |A^{\langle i \rangle}_{v_x v_y})|}.
\end{equation}

Here, $A$ is the adjacency matrix and $\beta$ is a free parameter ($\beta>0$) that is also called a ``damping factor''. One can realize that KI yields to a very similar score when $\beta$ is low enough as the paths that have higher lengths contribute less, and the similarity index is simply determined by the shorter paths~\cite{lu2011link}.  

In the case of $\beta < \frac{1}{\lambda_1^A}$, where $\lambda_1^A$ is the largest eigenvalue of the adjacency matrix, the similarity matrix can be written, as follows:
\begin{equation}
   S^{KI} = (I-\beta A)^{-1}-I,
\end{equation}
where $I$ is the identity matrix. 

\subsubsection{Global Leicht-Holme-Newman Index (GLHN)} 
The idea behind GLHN is very similar to that of KI, since it also considers a high similarity for the nodes if the number of paths between these corresponding nodes is high \cite{leicht2006vertex}, such that:
\begin{equation}
    S^{GLHN} = \beta_1(I-\beta_2 A)^{-1},
\end{equation}
where $\beta_1$ and $\beta_2$ are free parameters, and a smaller value of $\beta_2$ considers higher importance for the shorter paths, and vice versa. 

\subsubsection{SimRank (SR)} This index computes the similarity starting from the hypothesis ``two objects are similar if they are related to similar objects'', and it is recursively defined \cite{jeh2002simrank}. SR is equal to 1 when node \mbox{$v_x=v_y$, otherwise:}
\begin{equation}
  s_{(v_x,v_y)}^{SR} = \gamma . \frac{\sum_{v_{z_1} \in \Gamma(v_x)} \sum_{v_{z_2} \in \Gamma(v_y)} s_{(v_{z_1},v_{z_2})}^{SR}}{|\Gamma(v_x)|. |\Gamma(v_y)|},
\end{equation}
where $\gamma\in [0,1] $ is called decay factor and it controls how fast the effect of neighbor node pairs ($v_{z_1}$ and $v_{z_2}$) reduces as they move away from the original node pairs ($v_x$,$v_y$). SR can be explained in terms of a random walk process, which is, $s_{(v_x,v_y)}^{SR}$ measures how long the two random walkers are expected to meet on a particular node, starting with the $v_x$ and $v_y$ nodes. Its applicability is constrained on large networks due to its computational complexity \cite{wang2015link,liben2005algorithmic}.

\subsubsection{Pseudo-inverse of the Laplacian Matrix (PLM)}
Using Laplacian matrix 
$L=D-A$ rather than Adjacency matrix $A$ gives an alternative representation of a graph, where $D$ is the diagonal matrix of vertex degrees \cite{spielman2007spectral} ($D_{i,j}=0$ and $D_{i,i}=\sum_j{A_{i,j}}$). The Moore--Penrose pseudo-inverse of the Laplacian matrix, represented by $L^+$, can be used in the calculation of proximity measures \cite{fouss2007random}. Because PLM is calculated as inner product cosine similarity, it is also called "cosine similarity time" in the literature \cite{wang2015link}, and can be calculated as: 
\begin{equation}
    s_{(v_x,v_y)}^{PLM} = \frac{L_{(v_x,v_y)}^+}{\sqrt{L_{(v_x,v_x)}^+L_{(v_y,v_y)}^+}}.
\end{equation}

\subsubsection{Hitting Time (HT) and Average Commute Time (ACT)}
Motivated by random walk, as introduced by mathematician Karl Pearson \cite{pearson1905problem}, HT is defined as the average number of steps to be taken by a random walker starting from $v_x$ to reach node $v_y$. Because HT is not a symmetric metric, one may consider using ACT, which is defined as the average number of steps to be taken by the random walker starting from $v_x$ to reach the node $v_y$, and that from $v_y$ to reach node $v_x$. Therefore, HT can be computed by:
\begin{equation}
    s_{(v_x,v_y)}^{HT} = 1 + \sum_{v_z \in \Gamma(v_x)}{P_{v_x,v_z} s_{(v_z,v_y)}^{HT}}.
\end{equation}

Here, $P_{i,j}=D^{-1}A$, where $A$ and $D$ are the adjacency and the diagonal matrix of vertex degrees~\cite{wang2015link}. Accordingly, ACT can be formulated as:
\begin{equation}
    s_{(v_x,v_y)}^{ACT} = s_{(v_x,v_y)}^{HT} + s_{(v_y,v_x)}^{HT}.    
\end{equation}

For the sake of computational simplicity, ACT can be computed in a closed form using the pseudo-inverse of the Laplacian matrix of the graph, as follows \cite{fouss2007random}:
\begin{equation}
    s_{(v_x,v_y)}^{ACT} = m(L_{(v_x,v_x)}^+ + L_{(v_y,v_y)}^+ - 2L_{(v_x,v_y)}^+).
\end{equation}

One challenge of HT and ACT is that it gives very small proximity measures when the terminal node has high stationary probability $\pi_{v_y}$, regardless of the identity of the starting node. This~problem can be solved by normalizing the scores as $-s_{(v_x,v_y)}^{HT}.\pi_{v_y}$ and  $-(s_{(v_x,v_y)}^{HT}.\pi_{v_y} +s_{(v_y,v_x)}^{HT}.\pi_{v_x}$), respectively~\cite{liben2007link}. 

\subsubsection{Rooted PageRank (RPR)} 
PageRank (PR) is the metric that is used by Google Search in order to determine the relative importance of the webpages by treating links as a vote. Motivated by PR, RPR defines that the rank of a node is proportional to the likelihood that it can be reached through a  random walk \cite{wang2015link}, such that: 
\begin{equation}
    s_{(v_x,v_y)}^{RPR} = (1-\beta)(1-\beta P_{v_x,v_y})^{-1}.
\end{equation}

Here, $P_{i,j}=D^{-1}A$, where $A$ is the adjacency matrix and $D$ is the diagonal matrix of vertex degrees. It should be noted that one can calculate the PR by averaging the columns of RPR \cite{song2009scalable}.

\subsubsection{Escape Probability (EP)} The metric, which can be derived from RPR, measures the likelihood that the random walk starting from node $v_x$ visits node $v_y$ before coming back to the node $v_x$ again \cite{tong2007fast}. Let $Q(v_x,v_y)$ be equal to $(1-\beta D^{-1}A)^{-1}=s_{(v_x,v_y)}^{RPR}/(1-\beta)$; the equation of EP can be written, as follows \cite{song2009scalable}:
\begin{equation}
    s_{(v_x,v_y)}^{EP}=\frac{Q(v_x,v_y)}{Q(v_x,v_x).Q(v_y,v_y)-Q(v_x,v_y).Q(v_y,v_x)}.
\end{equation}

\subsubsection{Random Walk with Restart (RWR)} In a random walk (RW) algorithm, the probability vector of reaching a node starting from the node $v_x$ can be defined as:
\begin{equation}
    \vec{p_{v_x}}(t) = M^T\vec{p_{v_x}}(t-1),
\end{equation}
where $M$ is called the transition probability matrix, and it can be calculated by $A_{i,j}/ \sum_{k}{A_{i,k}}$, where $A$ is the adjacency matrix \cite{lu2010link}. Because RW does not yield a symmetric matrix, the metric of RWR, very similar to RPR, looks for the probability that a random walker starting from node $v_x$ visits node $v_y$ and comes back to the initial state node $v_x$ at the steady-state, such that:
\begin{equation}
    s_{(v_x,v_y)}^{RW} = \vec{{p_{v_x}}}^{v_y} + \vec{{p_{v_y}}}^{v_x}.
\end{equation}

\subsubsection{Maximal Entropy Random Walk (MERW)} 
The basic MERW algorithm, which is based on the maximum uncertainty principle, was proposed as a result of the necessity in order to define uniform path distribution in Monte Carlo simulations \cite{hetherington1984observations}. However, its applications on the stochastic models are very recent \cite{duda2012extended}. Li et al. \cite{li2011link} proposed MERW, which maximizes the entropy of a random walk, as follows:
\begin{equation}
    \lim_{l \rightarrow \infty} \frac{-\sum_{A^t_{v_x v_y} \in A^t} p(A^t_{v_x v_y}) \ln{p(A^t_{v_x v_y})}}{t}.
\end{equation}

Here, $p(A^t_{v_x v_y})$ is the multiplication of the iterative transition matrices ($M_{v_x v_z}.M_{v_z v_q}...M_{v_q v_y}$), where $M_{ij}$ can be calculated, as follows:
\begin{equation}
    M_{v_i v_j}=\frac{A_{v_i v_j}}{\lambda}\frac{\psi_{v_j}}{\psi_{v_i}},
\end{equation}
\textls[-5]{where $A$ is the adjacency matrix and $\psi$ is the normalized eigenvector with normalization constant $\lambda$ \cite{martinez2017survey}.}

\subsubsection{The Blondel Index (BI)} The index is proposed by Blondel et al. \cite{blondel2004measure} in order to measure the similarity for the automatic extraction of synonyms in a monolingual dictionary. Although BI is used to quantify the similarity between two different graphs, Martinez et al. show that investigating the similarity of two vertices in a single graph can also be evaluated in an iterative manner, as:   
\enlargethispage{0.5cm}
\begin{equation}
    S(t) = \frac{A S(t-1) A^T + A^T S(t-1)A}{||A S(t-1)A^T+A^TS(t-1)A||_F},    
\end{equation}
where \emph{S}(\emph{t}) refers to the similarity matrix in iteration t and $S(0) = I$. $||M||_F$ is the Frobenius matrix norm and it can be calculated, as follows: 
\begin{equation}
    ||M_{m \times n}||_F = \sqrt{\sum_{i=1}^m \sum_{j=1}^n {M_{i,j}^2}}.
\end{equation}

The similarity metric is obtained when \emph{S}(\emph{t}) is converged, such that $s_{(v_x,v_y)}^{BI}=S_{v_x,v_y}(t=c)$, where~$t=c$ denotes the steady state level. 

\subsection{Quasi-Local Similarity-Based Approaches}

The trade-off between the efficiency of the information regarding the whole network topological structure for the global approaches and the less time complex algorithms for the local-based methods have resulted in the emergence of quasi-local similarity-based methods for link prediction. Similarly,~these approaches are limited in the calculation of the similarity between arbitrary node pairs. However, quasi-local similarity methods provide an opportunity for computing the similarity between a node and the neighbors of its neighbors. Although some of the quasi-local similarity-based methods consider the whole topology of the network, their time complexity is less than that of global similarity-based approaches. 

\subsubsection{The Local Path Index (LPI)}  
The index, which is very similar to the well-known approaches KI and CN, considers the local path with a wider perspective by not only employing the information of the nearest neighbors, but~also the next two and three nearest neighbors \cite{lu2009similarity,zhou2009predicting}, such that:
\begin{equation}
    S^{LP} = A^2 + \beta A^3, 
\end{equation}
where $A$ is the adjacency matrix, $\beta$ is a free parameter to adjust the relative importance of the neighbors within the length $l=2$ distances and length $l=3$ distances. The metric can also be extended for the higher orders as:
\begin{equation}
    S^{LP^{(L)}} = \sum_{l=2}^{L}\beta^{l-2}{A^l}.
\end{equation}

The neighbors within the length of three distances are preferable due to increasing complexity in the higher orders of LP. One can easily realize that this similarity matrix simplifies to CN when $l=2$ and may produce a very similar result to KI given low $\beta$ values without the inverse transform process. The similarity between two nodes can be evaluated via $s_{(v_x,v_y)}^{LP}=S^{LP}_{v_x,v_y}$.

\subsubsection{Local (LRW) and Superposed Random Walks (SRW)}  Although the random walk-based algorithms perform well, the sparsity and computational complexity regarding massive networks are challenging for these algorithms. Thus, Liu and L\"{u} proposed the LRW metric \cite{liu2010link}, in which the initial resources for the random walker are assigned based on their importance in the graph. LRW considers the node degree as an important feature and it does not concentrate on the stationary state. Instead, the number of iterations is fixed in order to perform a few-step random walk. LRW can be formulated, as:
\begin{equation}
        s_{(v_x,v_y)}^{LRW}(t_{c})=\frac{|\Gamma(v_x)|}{2|E|}\vec{p^{v_x}_{v_y}}(t_{c})+\frac{|\Gamma(v_y)|}{2|E|}\vec{p^{v_y}_{v_x}}(t_c).
\end{equation}

Because superposing all of the random walkers starting from the same nodes may help to prevent the sensitive dependency of LRW to the farther neighboring nodes, SRW is proposed as:
\enlargethispage{0.5cm}
\begin{equation}
     s_{(v_x,v_y)}^{SRW}(t_{c}) = \sum_{t=1}^{t_{c}} s_{(v_x,v_y)}^{LRW}(t).
\end{equation}
\vspace{12pt}
\subsubsection{Third-Order Resource Allocation Based on Common Neighbor Interactions (RACN)} 
Motivated by the RA index, Zhang et al. \cite{zhang2014link} proposed RACN, in which the resources of nodes are allocated to the neighbors as:
\begin{equation}
    s_{(v_x,v_y)}^{RACN} = \sum_{v_z \in \Gamma(v_x) \cap \Gamma(v_y)} {\frac{1}{|\Gamma(v_z)|}} +\\ \sum_{e_{v_i,v_j} \in E,  |\Gamma(v_j)|<| \Gamma(v_i)|} (\frac{1}{|\Gamma(v_i)|}-\frac{1}{|\Gamma(v_j)|}),
\end{equation}
where $v_i \in \Gamma(v_x)$ and   $v_j \in \Gamma(v_j)$. The superiority of the RACN over the original RA has been shown in~\cite{zhang2018link} while using varying datasets.

\subsubsection{FriendLink Index (FL)} 
\textls[-5]{The similarity of two nodes is determined according to the normalized counts of the existing paths among the corresponding nodes with varying length $L$. The formulation for the FL index is as follows:}
\begin{equation}
    s_{(v_x,v_y)}^{FL}=\sum_{l=1}^{L}{\frac{1}{l-1}\frac{|A^l_{v_x,v_y}|}{\prod_{j=2}^l{(|V|-j)}}},
\end{equation}
where $|V|$ is the number of vertices in the graph. The metric is favorable, due to its high performance and speed \cite{papadimitriou2012fast}.

\subsubsection{PropFlow Predictor Index (PFP)} 
PFP is a metric that is inspired by Rooted PageRank, and it simply equals the probability that the success of random walk starts from node $v_x$ and terminates at node $v_y$ in not more than $l$ steps \cite{lichtenwalter2010new}. This restricted random walk selects the links based on weights, denoted as $\omega$ \cite{wang2015link}, such that: 
\begin{equation}
    s_{(v_x,v_y)}^{PFP}=s_{(v_a,v_x)}^{PFP}\frac{\omega_{v_x v_y}}{\sum_{v_z \in \Gamma(v_x)}{\omega_{v_x v_y}}}.
\end{equation}

The most important superiority of PFP is its widespread use in directed, undirected, weighted, unweighted, sparse, or dense networks.

\section{Probabilistic Methods}
Probabilistic models are supervised models that use Bayes rules. The most important drawback of some of these models is their being slow and costly for large networks \cite{al2011survey}. In the following, we~introduce the five most important probabilistic methods of link prediction.  
\subsection{Hierarchical Structure Model}
\label{sec: hierarchical}

This model was developed based on the observation that many real networks present a hierarchical topology \cite{ravasz2003hierarchical}. This maximum likelihood-based method searches for a set of hierarchical representations of the network and then sorts the probable node pairs by averaging over all ofnthe hierarchical representations explored. The model was first proposed in the work of \cite{clauset2008hierarchical}, in which it develops a hierarchical network model that can be represented by a dendrogram, with $|N|$ leaves and $|N-1|$ internal nodes. Each leaf is a node from the original network and each internal node represents the relationship of the descendent nodes in the dendrogram. A value of $p_{r}$ is also attributed to each internal node $r$, which represents the probability with which a link exists between the branches descending from it. If $D$ is a dendrogram that represents the network, the likelihood of dendrogram with a set of internal node probabilities ($p_{r}$) is:
\begin{equation}
\label{eq: hierarchicalmodel}
    \mathcal{L}\left ( D,\left \{ p_{r} \right \} \right )= \prod_{r \in D} p_{r}^{E_{r}} (1-p_{r})^{L_{r}R_{r}-E_{r}}.
\end{equation}

In the above equation, $E_{r}$ is the number of links that connect nodes that have a node $r$ as their lowest common ancestor in D. $L_{r}$ and $R_{r}$ represent the number of leaves in the left and right subtrees that are rotted in r, respectively. Setting $p_r^* = \frac{E_{r}}{L_{r}R_{r}}$ maximizes the likelihood function~(\ref{eq: hierarchicalmodel}). Replacing $p_{r}$ with $p_r^*$ in the likelihood function, the likelihood of a dendrogram at its maximum can be calculated~by:
\begin{equation}
    \mathcal{L}(D) = \prod_{r \in D} \left [(1-p_r^*)^{1-p_r^*}  p_r^{*^{p_r^*}}  \right ]^{L_{r} R_{r}}.
\end{equation}

These equations are then utilized to perform link prediction. After a Markov Chain Monte Carlo method is used to sample a large number of dendrograms with probabilities proportional to their likelihood, the connection probability between two nodes $v_i$ and $v_j$ is estimated by averaging over all the sampled dendrograms. This task is performed for all sampled dendrograms and, subsequently, the~node pairs are sorted based on the corresponding average probabilities. The higher the ranking, the~more likely that the link between the node pair exists. A major drawback of the hierarchical structural model is its computational cost and being very slow for a network consisting of a large set of nodes.  

\subsection{Stochastic Blockmodel}
\label{sec: stochasticblockmodel}

Stochastic block models are based on the idea that nodes that are heavily interconnected should form a block or community \cite{goldenberg2010survey}. In a stochastic block model, nodes are separated into groups and the probability that two nodes are connected to each other is merely dependent on the group to which they belong \cite{guimera2009missing}. Stochastic block models have been successfully applied to model the structure of complex networks \cite{peixoto2014hierarchical,valles2018consistencies}. They have also been utilized to predict the behavior in drug interactions~\cite{guimera2013network}. The~work of \cite{rovira2013predicting} uses a block model in order to predict conflict between team members. \cite{godoy2016accurate} also utilizes a stochastic block model in order to develop a probabilistic recommender system. 

As noted above, the probability that two nodes $i$ and $j$ are connected depends on the blocks that they belong to. A block model $M=(P, Q)$ is completely determined by the partition $P$ of nodes into groups and the matrix $Q$ of probabilities of linkage between groups. While numerous partitions (models) can be considered for a network, the likelihood of a model $A^O$ can be calculated by the following \cite{holland1983stochastic,guimera2009missing}:
\begin{equation}
\label{eq: blockmodel}
    \mathcal{L}(A^O|P,Q) = \prod_{\alpha \leq \beta} Q_{\alpha \beta}^{l_{\alpha \beta}^O} (1-Q_{\alpha \beta })^{r_{\alpha \beta}-l_{\alpha \beta}^O}.
\end{equation}

In Equation~(\ref{eq: blockmodel}), $l_{\alpha \beta}^O$ is the number of links in $A^O$ between nodes in groups $\alpha$ and $\beta$ of $P$, and $r_{\alpha \beta}$ is the maximum number of links possible, which is $|\alpha||\beta|$ when $\alpha \neq \beta$ and $|\alpha|\choose 2$ when $\alpha=\beta$ . Setting $Q_{\alpha \beta}^* = \frac{l_{\alpha \beta}^O}{r_{\alpha \beta}}$ maximizes the likelihood function~(\ref{eq: blockmodel}). By applying Bayes theorem, the probability (reliability) of a link with maximum likelihood can be computed. 

\begin{figure}[H]
\centering
		\includegraphics[width=1\linewidth, keepaspectratio]{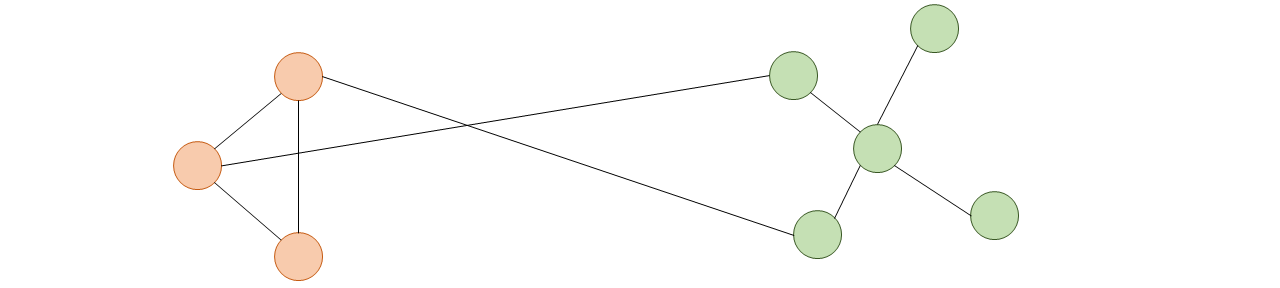}
		\caption{{An example of Blockmodel likelihood calculation Here, a probable partitioning is presented}. The block on left is $\alpha$ and the block on right is $\beta$. $Q_{\alpha\alpha}^*=1$, $Q_{\alpha\beta}^*=\frac{2}{12}$, and $Q_{\beta\beta}^*=\frac{5}{6}$. Hence, likelihood is calculated, as follows: $1^3\times1\times\frac{2}{15}^2\times\frac{13}{15}^{13}\times\frac{1}{2}^5\times\frac{1}{2}^5\approx2.701\times10^{-6}$.}
		\label{fig:relativeActivity}
	\end{figure}

Similar to the hierarchical structure model that is discussed in Section~\ref{sec: hierarchical}, a significant shortcoming of this method is that it is very time-consuming. While the Metropolis algorithm \cite{bhanot1988metropolis} can be utilized to sample partitions, this approach is still impractical for a large network.  An example of blockmodel likelihood calculation is illustrate in Figure~\ref{fig:relativeActivity}.

\subsection{Network Evolution Model}
Ref. \cite{kashima2006parameterized} proposed a network topology based model for link prediction. In this model, probabilistic flips of the existence of edges are modeled by a ``copy-and-paste'' process between the edges \cite{kashima2006parameterized}. The~problem of link prediction is defined, as follows: the data domain is represented as a graph $G=(V,s)$, where $V$ is the set of nodes of the network and $s: V \times V \rightarrow [0,1]$ is an edge label function. $s(v_i,v_j)$ indicates the probability that an edge exists between $i$ and $j$. $s^{(t)}$ shows the edge label function at time $t$, and its of Markovian nature, i.e., $s^{(t+1)}$ only depends on $s^{(t)}$. The fundamental idea behind the proposed network edge label copy-and-paste mechanism is that, if a node has a strong influence on another node, the second nodes association will be highly affected by the second node. The probability of an edge existing between nodes $i$ and $j$ at time $t+1$ is as follows:
\begin{equation}
\begin{split}
\label{eq: networkevolutionmodel}
s^{t+1}(v_i,v_j) = &\frac{1}{|V|-1}(\sum_{k\neq i,j} w_{v_k v_j}s^{(t)} (v_k,v_i)+ w_{v_k v_i}s^{(t)} (v_k,v_j)) + \\ &(1-\frac{1}{|V|-1}\sum_{k\neq i,j}w_{v_k v_j}+w_{v_k v_i})s^{(t)}(v_i,v_j),
\end{split}
\end{equation}
where $w_{v_k v_j}$ is the probability that an edge label is copied from node $v_k$ to node $v_j$. In Equation~(\ref{eq: networkevolutionmodel}), the first term represents the probability that the edge label for $(v_i,v_j)$ is changed by copy and pasting. The second term represents when the same edge label is unchanged. The linkages are obtained by iteratively updating Equation~(\ref{eq: networkevolutionmodel}) until convergence. The objective function according to which the parameters are set is solved by an expectation maximization type transductive learning. 

\subsection{Local Probabilistic Model}
The work of \cite{wang2007local} proposed a local probabilistic model for link prediction, in which the focus of the original paper is particularly in the context of evolving co-authorship networks. Given the candidate link, e.g. nodes $v_i$ and $v_j$, first, the central neighborhood set of $v_i$ and $v_j$ are determined, which is the set of nodes that are the most relevant to estimating the co-occurrence probability. The~central neighborhood sets are chosen from the nodes that lie along paths of shorter length between $v_i$ and $v_j$. Ref. \cite{wang2007local} proposes an algorithm in order to determine central neighborhood set, which is, as follows: first, collecting all of the nodes that lie on length-2 simple paths, then those on length-3 simple paths, and so on. The paths are then ordered based on the frequency scores and the ones with the highest scores are chosen \cite{wang2007local}. A path length threshold is also considered for the sake of decreasing computational cost (\cite{wang2007local} proposes a threshold of 4 for their specific problem). Next, they form a transaction dataset that is formed by a chronological set of events (co-authoring articles). A~non-derivable itemset mining is performed on this dataset, which results in all non-redundant itemsets along with their frequencies. In the end, a Markov Random Field (MRF) graph model is trained while using the derived dataset. The resulting final model gives the probability of the existence of each link $v_i$ and $v_j$. 

\subsection{Probabilistic Model of Generalized Clustering Coefficient}
\begin{figure}[H]
\centering
		\includegraphics[height=0.25\linewidth, keepaspectratio]{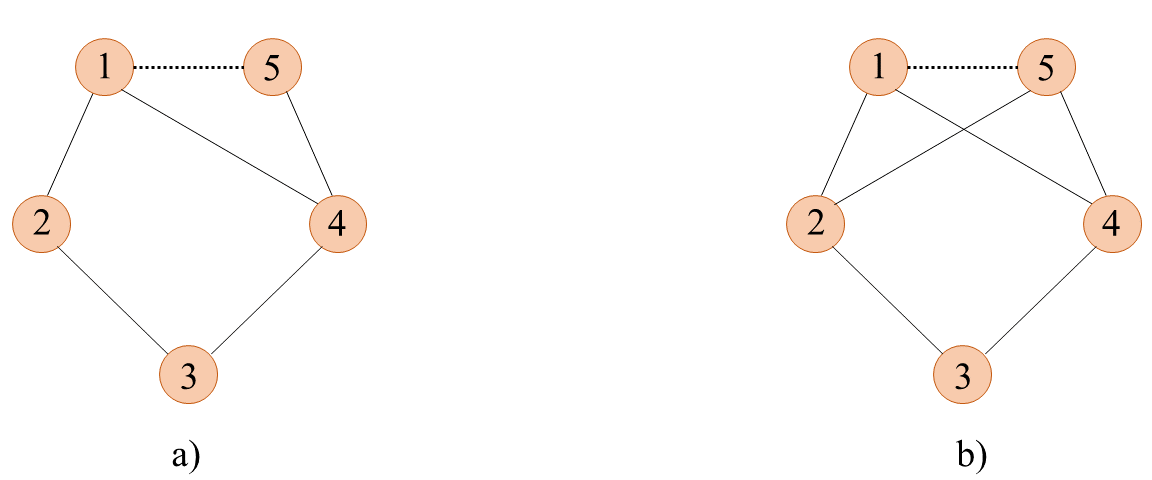}
		\caption{ (\textbf{a}) There is only one path of length two from node $1$ to node $5$ in the network and $\textnormal{Pr}_2((1,5) \in E)=c_{2}$ . (\textbf{b}) There are two paths of length 2 from the node $1$ to node $5$, therefore $\textnormal{Pr}_2((i,j) \in E)=\frac{c_{2}^2}{c_{2}^2+(1-c_2)^2} $.
		}
		\label{fig:cycle}
	\end{figure}

This method that was proposed by \cite{huang2010link} focuses on analyzing the predictive power of clustering coefficient \cite{huang2010link}. The generalized clustering coefficient $C(k)$ of degree $k$ is defined as \cite{huang2010link}:
\begin{equation}
\label{eq: cyclicprobability}
C(k)=\frac{\textnormal{number of cycles of length \textit{k} in the graph}}{\textnormal{number of paths of length \textit{k}}}
\end{equation}
As explained in \cite{huang2010link}, generalized clustering coefficients describe the correlation between cycles and paths in a network. Therefore, the probability of formation of a particular link is determined by the number of cycles (of different lengths) that will be constructed by adding that link \cite{huang2010link}. The concept of cycle formation model is explained, as follows: a cycle formation model of degree $k$ $(k\geq1)$ is governed by k link generation mechanisms, $g(1)$, $g(2)$,..., $g(k)$, which are each described by $c_1$, $c_2$,..., $c_k$. If $P_{v_i v_j k}$ shows a path from $v_i$ to $v_j$ with length $k$, then $c_k=P((v_i,v_j)\in E||P_{v_i v_j k}|=1)$ (the probability that there is a link between $i$ and $j$, given that there is one path of length $k$ between them). We know that, if there is more than one path with length $k$ from $v_i$ to $v_j$, then the probability that there is a link between them increases (see Figure~\ref{fig:cycle} for instance). Therefore:
\begin{equation}
\label{eq: cyclicprobability2}
\begin{split}
\textnormal{if } &\textnormal{P}((v_i,v_j) \in E||P_{v_i,v_j,k}|=1)=c_{k}\textnormal{    } \& \textnormal{    } |P_{v_i,v_j,k}|=m \textnormal{ }\rightarrow \\ &\textnormal{P}((v_i,v_j) \in E||P_{v_i,v_j,k}|=m)= \frac{c_{k}^m}{c_{k}^m+(1-c_k)^m} 
\end{split}
\end{equation}

Because of the fact that the total link occurrence probability between $v_i$ and $v_j$ is a result of the effect of multiple mechanisms of cycle formation model of degree $k$ $(CF(k))$ is calculated by:
\begin{equation}
\label{eq: cyclicprobability3}
\begin{split}
&P_{m_2,...,m_k}=\textnormal{P}((v_i,v_j) \in E||P_{v_i v_j 2}|=m_2,...,|P_{v_i v_j k}|=m_{k}) = \\
&\frac{c_1c_2^{m_2}...c_k^{m_k}}{(c_1c_2^{m_2}...c_k^{m_k})+(1-c_1)(1-c_2)^{m_2}...(1-c_k)^{m_k}} 
\end{split}
\end{equation}

\section{Relational Models}
\label{sec: relationalmodels}

\begin{figure}[H]
		\includegraphics[width=1\linewidth, keepaspectratio]{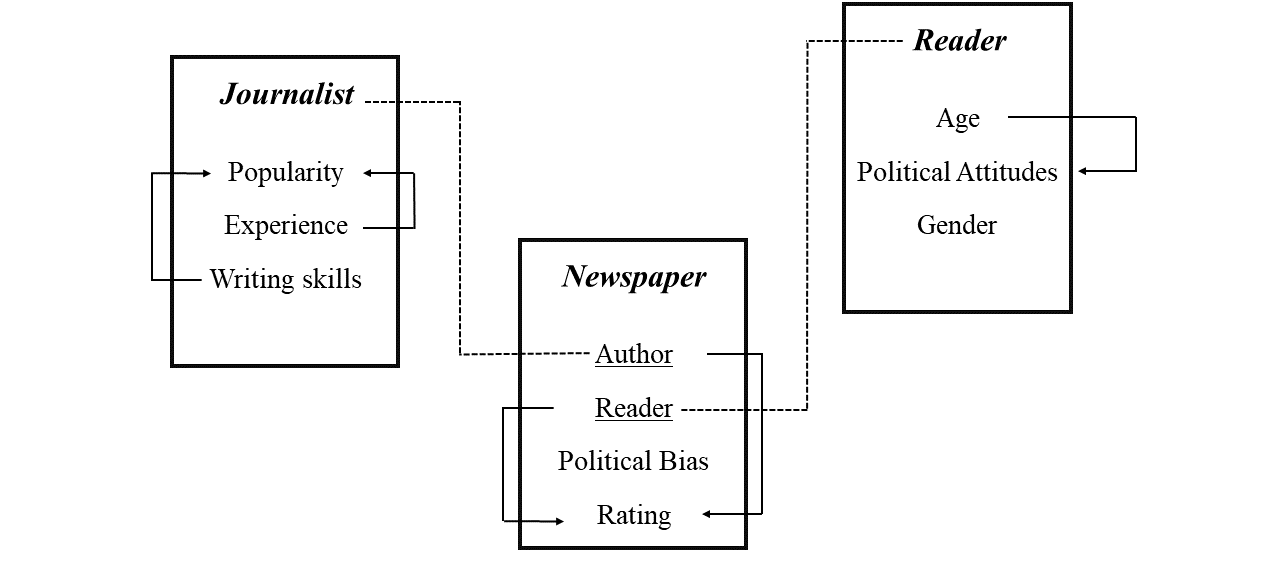}
		\caption{An example of a relational schema for a simple domain. The underlined attributes are reference slots of the class and the arrows show the types of objects to which they are referring. 
		}
		\label{fig:relational}
	\end{figure}

One drawback of the previously mentioned methods is that they do not incorporate vertex and edge attributes to model the joint probability distribution of entities and links that associate them~\cite{al2011survey}. Probabilistic Relational Models (PRM) \cite{friedman1999learning} is an attempt to use the rich logical structure of the underlying data that is crucial for complicated problems. One major limitation of Bayesian networks is the lack of the concept of an ``object'' \cite{getoor2007probabilistic}. Bayesian PRMs \cite{heckerman1995learning,friedman1999learning} include the concept of an object in the context of Bayesian networks, in which each object can have their attributes and relations exist between objects and their attributes. Figure~\ref{fig:relational} is an example of a schema for a simple domain. A relational model consists of a set of \emph{classes}, $\Upsilon=\{Y_1, Y_2,..., Y_n\}$. In Figure~\ref{fig:relational}, $\Upsilon=\{\textnormal{Journalist}, \textnormal{Newspaper}, \textnormal{Reader}\}$. Each class also contains some descriptive attributes, the set of which is shown with $A(Y)$. For example, Journalist has attributes Popularity, Experience, and Writing skills. In order for objects to be able to refer to other objects, each class is also associated with a set of reference slots, which is shown by $Y.\rho$. Slot chains also exist, which are references between multiple objects (similar to $f(g(x))$). $Pa(Y.A)$ shows the set of parents of $Y.A$. For instance, in Figure~\ref{fig:relational}, a journalist's Popularity depends on her Experience and Writing skills. Dependency can also be a result of a slot chain, meaning that some attributes of a class depend on some attributes of another class. The joint probability distribution in a PRM can be calculated, as follows \cite{getoor2007probabilistic}:
\begin{equation}
\label{eq: PRM}
P(I|\sigma_r,S,\theta_S) =  \prod_{Y_i} \prod_{A \in A(Y_i)} \prod_{y \in \sigma_r (Y_i)} P(I_{y.A}|I_{Pa(y.A)})
\end{equation}

In Equation~(\ref{eq: PRM}), $I$ shows an instance of a schema $S$, which specifies for each class $Y$, the set of objects in the class, a value for each attribute $y.A$, and a value for each reference slot $y.\rho$. Additionally, $\sigma_r$ is a relational skeleton, which denotes a partial specification of an instance of a schema, and it specifies the set of objects for each class and the relations that hold between the objects \cite{getoor2007probabilistic}.

The task of link prediction can then be performed by considering the probability of the existence of a link between two objects in the relational model \cite{getoor2002learning}. The work of \cite{bilgic2007combining} shows that deriving the distribution of missing descriptive attributes will benefit from the estimation of link existence likelihood. Besides, a Relational Bayesian Network, in which the model graph is a directed acyclic graph, the Relational Markov Network is also proposed \cite{taskar2012discriminative,taskar2004link}, in which the graph model is an undirected graph and it can be utilized for the task of link prediction. Relational Markov Networks address two shortcomings of directed models: They do not constrain the graph to be acyclic, which allows for various possible graph representations. Additionally, they are well suited for discriminative training \cite{taskar2007relational}.

There exist other relational models for the task of link prediction. The DAPER model is a directed acyclic type of probabilistic entity-relationship model \cite{heckerman2007probabilistic}. The advantage of the DAPER model is being more expressive than the aforementioned models \cite{heckerman2004probabilistic}. Other Bayesian relational models in the literature include stochastic relational model \cite{yu2007stochastic}, which models the stochastic structure of entity relationships by a tensor of multiple Gaussian processes \cite{lu2011link}, relational dependency network \cite{neville2007relational,heckerman2013dependency}, and parametric hierarchical Bayesian relational model \cite{xu2005dirichlet}.

\section{Learning-Based Methods}
The feature extraction-based methods that are discussed earlier in this paper provide a starting point for the systematic prediction of missing or future associations available through learning the effective attributes. Among these effective features for link prediction, employing the topological attributes that can be extracted from the graph structure is the foundation of all learning-based link prediction algorithms, from which the pair-wise shortest distance attribute is the most common topological feature. Besides the topological attributes, some machine learning models benefit from the node and domain specific attributes, referred to as the aggregated and proximity features, respectively~\cite{al2006link}. 

The introduction of supervised learning algorithms to the problem of link prediction led to the state-of-the-art models that achieve high prediction performances \cite{duan2017ensemble}. These models view the problem of link prediction as a classification task. In order to approach the link prediction problem, supervised models are supposed to tackle a few challenges, including the unbalanced data classes that result from the sparsity property of real networks, and the extraction of the topological, proximity, and aggregated attributes as independent informative features \cite{hamilton2017representation}. 
There is extensive literature on the classification models for link prediction, including the application of traditional machine learning methods into this field of research. Support Vector Machines, K-nearest Neighbors, Logistic Regression, Ensemble Learning, and Random Forrest, Multilayer Perceptron, Radial Basis Function network, and Naive Bayes are just a few of the supervised learning methods that are extensively used in link prediction. A comparison between a few of these supervised methods has been presented in \cite{al2006link}, where, surprisingly, SVM with RBF kernel is reported to be very successful in the accuracy and low squared error of the model. 

Although the traditional machine learning models for link prediction rely on user-defined feature encoding, the evolution of these models has led to the generation of automatic feature encoders, which prevent hand-engineered attributes \cite{hamilton2017representation}. These models aim to learn graph encoding, node, and/or domain-related features into low-dimensional space, and are referred to as representation learning or graph embedding-based models for link prediction. These methods can be trained while using neural networks or dimensionality reduction algorithms \cite{grover2016node2vec}. The applications of graph analysis and representation learning has led to the development of advanced language models that focus on language understanding, relation discovery, and question answering. Knowledge graphs, which represent sequences of relations between named entities within a textual content, are being widely investigated for the task of link prediction, relation prediction, and knowledge graph completion \cite{liu2020k}. Although many of the reviewed methods in this survey are applicable to different applications and graph types, knowledge graphs and their embedding methods are dependent to directed relationships. Examples of recent methods for knowledge graphs are Relational Graph Convolutional Neural Networks (R-GCN) \cite{schlichtkrull2018modeling}, which are able to extract features from a given data and, accordingly, generate a directed multigraph, label node types, and their relationships in the generated graph, and, finally, generate a latent knowledge-based representation that can be used for node classification as well as link prediction. Other language models, such as Bidirectional Encoder Representations from Transformers (BERT) \cite{devlin2018bert}, which use pre-trained language models, and their variations, including Knowledge Graph BERT (KG-BERT) \cite{yao2019kg} and Knowledge-enabled BERT (K-BERT) \cite{liu2020k}, can extract node and relation attributes for knowledge graph completion and link prediction \cite{heidaribots}. A comprehensive review on embedding methods that are designed for knowledge graphs is available in \cite{wang2017knowledge}. 

The tasks of vertex representation learning and vertex collocation profiling (VCP) for the purpose of topological link analysis and prediction were introduced in \cite{khosla2019node} and \cite{lichtenwalter2012vertex}, respectively. Comprehensive information on the surrounding local structure of embedded pairs of vertices $v_x$ and $v_y$ in terms of their common membership in all possible subgraphs of $n$ vertices over a set of $r$ relations is available from their VCP, written as $VCP_{x,y}^{n,r}$, and the VCP elements are closely related to isomorphic subgraphs. Thus, this method helps in the understanding of link formation mechanism from the nodes and graph representation.

\begin{figure}[H]
\centering
		\includegraphics[width=0.8\linewidth, keepaspectratio]{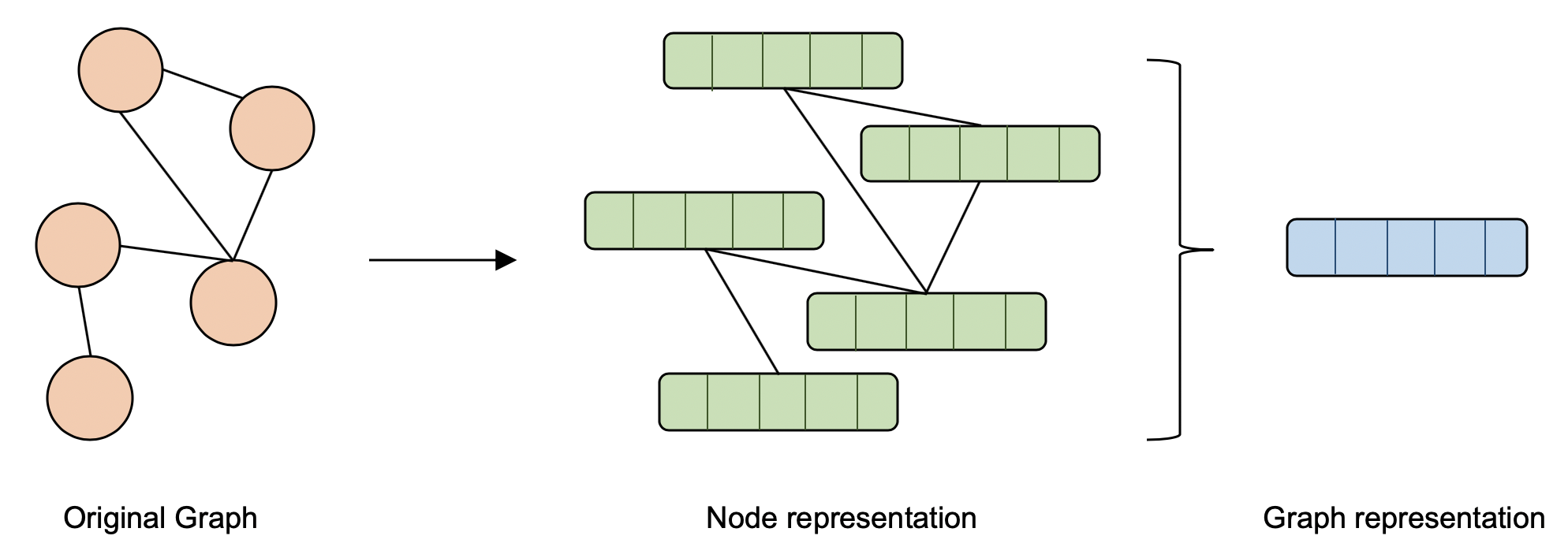}
		\caption{An example of node and graph representation. Here the node representation vectors are aggregated to generate a single graph representation.} 
		\label{fig:representation}
	\end{figure}

Mapping the graph to a vector space is also known as encoding. On the contrary, the reconstruction of the node neighborhood from the embedded graph is referred to as decoding. Graph representation can be learned via supervised or unsupervised methods while using an appropriate optimization algorithm in order to learn the embeddings  \cite{hamilton2017representation}. This mapping can be defined for graph G = $<$ V, E $>$ as \(f: v_x \rightarrow v_{x^\prime} \in \mathbb{R}^d, \forall x \in [n] \), such that \(d \ll \mid V\mid \), where $n$ denotes the total number of vertices, \(v_x\) is a sample node that has been embedded to \(d\)-dimensional vector space, and the embedded node is represented by $v_{x^\prime}$. Figure \ref{fig:representation} illustrates the procedure of node and graph representation.

Representation learning algorithms for the task of link prediction can be divided into categories based on their decoder function, a similarity measure for graphs, and the loss function in the models~\cite{hamilton2017representation}. Therefore, we categorize these methods into 
\begin{enumerate*}[label=(\roman*),leftmargin=21pt,labelsep=7pt]
\item Matrix Factorization-Based Models, 
\item Path and Walk-Based Models, and 
\item Deep Neural Network-Based Methods.
\end{enumerate*}

\subsection{Matrix Factorization-Based Methods}

These methods are able to extract latent features with additional graph features for link prediction. In these models, the vector representation of the topology-related features produces an N-dimensional space, where $N = |V|$ is the number of vertices in the network. 
The main purpose of matrix factorization-based methods is to reduce the dimensionality while also preserving the nonlinearity and locality of the graph via employing deterministic measures of node similarity in the graph. However, the global structure of the graph topology may be generally lost \cite{tang2015line}. 

SVD is one of the commonly used methods as a result of its feasibility in low-rank \mbox{approximations \cite{cui2017survey,ou2016asymmetric}}. Here, the link function $L(.)$ is defined as $G \approx L(U \Lambda U^T)$, where $U \in \mathbb{R}^{|V|\times k}, \Lambda \in \mathbb{R}^{k \times k}$, and $k$ denotes the number of latent variables in SVD. The similarity $s(v_x,v_y)$ between the node pairs $v_x$ and $v_y$ is defined by $L(u_{v_x}^T \Lambda u_{v_y})$.   

In \cite{menon2011link}, a latent feature learning method for link prediction has been proposed by defining a latent vector $\vv{l_{v_x}}$ and a feature vector $\vv{a_{v_x}}$ for each node $v_x$, a weight vector $W_{v}$ for node features, a weight vector $\vv{w_{e}}$ for edge features, and a vector of features $\vv{b_{v_x, v_y}}$ for each edge. This model computes the prediction of edge formation as:
\begin{equation}
    s(v_x, v_y) = \frac{1}{1+exp(-\vv{l_{v_x}}^T F\vv{l_{v_y}} - \vv{a_{v_x}}^T W_v\vv{a_{v_y}} - \vv{w_e}^T \vv{b_{v_x, v_y}})},
\end{equation}
where, $F$ is the scaling factor for each edge. 

The inner-product-based embedding models for link prediction embed the graph based on a pairwise inner-product decoder, such that the node relationship probability is proportional to the dot product of node embeddings:
\begin{equation}
    DEC(v_{x^\prime},v_{y^\prime}) = v_{x^\prime}^T v_{y^\prime},
\end{equation}
\begin{equation}
    \mathcal{L} = \sum_{(v_x,v_y) \in D}{|| DEC(v_{x^\prime},v_{y^\prime}) - s_G(v_x,v_y)||^{2}_2}.
\end{equation}

Graph Factorization (GF) \cite{ahmed2013distributed}, GraRep \cite{cao2015grarep}, and HOPE \cite{ou2016asymmetric} algorithms are examples of the inner-product-based methods for link prediction. Graph factorization model partitions the graph by minimizing the number of neighboring nodes, rather than applying edge cuts, as the storage and exchange of parameters for the latent variable models and their inference algorithms are related to nodes. 
HOPE \cite{ou2016asymmetric} focuses on the representation and modeling of directed graphs, as directed associations can represent any type of graph. This model preserves the asymmetric transitivity for directed graph embeddings. The asymmetric transitivity property captures the structure of the graph by keeping the correlation between the directed edges, such that the probability of the existence of a directed edge from $v_x$ to $v_y$ is high if a directed edge exists for the opposite direction. 
HOPE supports classical similarity measures as proximity measurements in the algorithm, including the Katz Index (KI), Rooted PageRank (RPR), Common Neighbors (CN), and Adamic-Adar (AA). 

\subsection{Path and Walk-Based Methods}

The developed models for link prediction that are designed based on random walk statistics prevent the need for any deterministic similarity measures. In these algorithms, similar embeddings are being produced for nodes that co-occur on graph short random walks. These algorithms investigate the node features, including node centrality and similarity via graphs exploration and sampling with random walks or search algorithms, such as Breadth First Search (BFS) and Depth First Search (DFS)~\cite{goyal2018graph}. The random walk-based models for graphs can be divided into many different categories, according to varying perspectives. One possible division for these models includes categorization that is based on their embedding output, for instance, local structure-preserving methods, global structure-preserving methods, and the combination of the two \cite{zhou2018graph}. 

Representations with BFS provide information regarding the similarity of nodes in the case of their roles in the network, for instance, representing a hub in the graph \cite{grover2016node2vec}. On the contrary, random walks with DFS can provide information regarding the communities that nodes belong to. These~algorithms have been recently applied along with generative models to introduce edges and nodes directly to the graph \cite{wang2017graphgan}. Community aware random walk for network embedding (CARE), as introduced in  \cite{keikha2018community}, is another approach for the task of link prediction and multi-label classification. This model builds customized paths that are based on local and global structures of network, and uses the Skip-gram model to learn representation vectors of nodes.

In comparison to walk-based methods, link prediction that is based on meta path similarity has been introduced in \cite{sun2011pathsim}, which operates a similarity search among the same type of nodes. Thus, meta path-based methods extend link prediction to heterogeneous networks with different types of vertices. In this model, a meta path refers to a sequence of relations between object types and defines a new composite relation between its starting type and ending type. The similarity measure between two objects can be defined according to random walks used in P-PageRank, pairwise random walk used in SimRank, P-PageRank, or SimRank on the extracted sub-graph or, finally, using PathSim, which captures the subtle semantics of similarity among peer objects \cite{sun2011pathsim}. PathSim calculates the similarity of two peer objects as:
\begin{equation}
    s(v_x, v_y) = \frac{2 \times |\{p_{v_x,v_y}:p_{v_x,v_y \in \mathcal{P}}\}|}{|\{p_{v_x,v_x}:p_{v_x,v_x \in \mathcal{P}}\}| + |\{p_{v_y,v_y}:p_{v_y,v_y \in \mathcal{P}}\}|},
\end{equation}
where $\mathcal{P}$ refers to the meta path defined on the graph of network schema, $p_{v_x, v_y}$ is a path instance between $v_x$ and $v_y$, and $p_{v_x, v_x}$ and $p_{v_y, v_y}$ are the same concept for vertices $v_x$ and $v_y$. An application of using meta-path for link prediction is in \cite{fu2016predicting}, which predicts drug target interactions (DTI) on the observed topological features of a semantic network in the context of drug discovery. 

\subsection{Neural Network-Based Methods}

In order to avoid strong assumptions for every heuristic related to node similarities and edge formation, link prediction algorithms that are based on neural networks have been proposed that automatically learn a suitable heuristic from a given network. In \cite{zhang2018link}, a mapping function for the subgraph patterns to link existence is being learned by extracting a local subgraph around each target link. Thus, this model automatically learns a ``heuristic'' that suits the graph. The powerful capabilities and simplicity of using neural network-based methods have led to the generation of a family of complex encoder-decoder-based representation learning models, such as Graph Neural Networks (GNNs) \cite{zhang2019heterogeneous,scarselli2008graph} and Graph Convolutional Neural Networks (GCNs) \cite{kipf2016semi,kipf2016variational,schlichtkrull2018modeling,berg2017graph}. 

Although the general concept of graph neural networks was first presented in \cite{scarselli2008graph}, many neural network-based algorithms for representation learning and link prediction have been proposed, including SEAL \cite{zhang2018link}, which uses GNNs to learn general graph structure features for link prediction from local enclosing subgraphs. Besides models that consider graph structure features, latent and explicit features are also investigated in the literature for link prediction. Furthermore, efficient strategies for capturing multi-modality for graphs, for instance, node heterogeneity, have been originated from neural network-based models. Another extension for graph embedding methods that have become achievable by neural networks, is the embedding of subgraphs $(S \subset V)$. The~attribute aggregation procedure in different neural network architectures may vary according to their connection types, and the usage of filters or gates in the propagation step of the models \cite{zhou2018graph}. 

In order to learn the information on the neighboring nodes, GNNs \cite{scarselli2008graph} aim to learn a state embedding $h_{v_x} \in \mathbb{R}^s$ iteratively, where s is the dimension for the vector representation of node $v_x$. By stacking the states for all of the nodes, the constructed vectors $H$, and the output labels $O$ can be represented as:
\begin{equation}
    H = F_g(H , X),
\end{equation}
\begin{equation}
    O = O_g(H , X_N),
\end{equation}
where $F_g$ is the global transition function, $O_g$ is the global output function, $X$ refers to the feature vector, and $X_N$ stands for the feature vector for all nodes. The updates per iteration can be defined as:
\begin{equation}
    H^{t+1} = F(H^t , X),
\end{equation}
where $t$ denotes the t\_th iteration. In this algorithm, the learning of the representations can be achieved by a supervised optimization method, such as the gradient-descent method. 

The SEAL \cite{zhou2018graph} algorithm that has been designed for the task of link prediction considers enclosing subgraph extraction for a set of sampled positive (observed) and negative links in order to prepare the training data for GNN and uses that information to predict edge formations. The GNN model receives the adjacency matrix ($A$) and node information matrix ($X$) as input, where each row of $X$ corresponds to a feature vector of a vertex. The process of $X$ preparation for each enclosing subgraph includes three components of structural node labels based on Double-Radius Node Labeling (DRNL), node embeddings, and node attributes. Another neural network-based model for the task of link prediction is HetGNN \cite{zhang2019heterogeneous}, which considers heterogeneous networks. This model starts with a random walk with restart strategy and samples a fixed size of correlated heterogeneous neighbors to group them based upon node types. Subsequently, neural network architecture with two modules is used in order to aggregate feature information of sampled neighboring vertices. The deep feature interactions of heterogeneous contents are captured by the first module, which generates content embedding for each vertex. The aggregation of content embeddings of different neighboring types is being done by the second module. HetGNN combines these outputs in order to obtain the final node embedding. 

Multi-layer Perceptrons (MLPs) are neural network-based representation learning algorithms that approach graph embedding via message passing, in which information flows from the neighboring nodes with arbitrary depth. Message Passing Neural Networks (MPNNs) \cite{gilmer2017neural} further extend GNNs and GCNs by proposing a single framework for variants of general approaches, such as incorporating the edge features in addition to the node features.

Graph Auto-Encoders (GAE) and Variational Graph Auto-Encoders (VGAE) \cite{kipf2016variational} are another category of graph neural networks that aim to learn the node representations in an unsupervised manner. The majority of models based on GAE and its derivations employ Graph Convolutional Networks (GCNs) for the node encoding procedure. Next, these algorithms employ a decoder in order to reconstruct the graph's adjacency matrix $A$. This procedure can be formally represented as:
\begin{equation}
    Z = GCN(X, A),
\end{equation}
where $Z$ is the convolved attribute matrix. GAEs can learn the graph structures while using deep neural network architectures, and reduce the graph dimensionality in accordance with the number of channels of the auto-encoder hidden layers \cite{cao2016deep}. Additionally, GAE-based models are able to embed the nodes into sequences with diverse lengths. This benefits the auto-encoders not only to achieve high performances for testing over the unseen node embeddings, but also to aggregate the node attributes in order to improve their prediction accuracy \cite{hamilton2017representation}. 
GC-MC \cite{berg2017graph} and Adversarially Regularized Graph Auto-Encoders (ARGA) are examples of representation models with auto-encoder architectures \cite{pan2018adversarially}. Auto-encoders are also being used without neural network architectures, for instance, LINE \cite{tang2015line}, DNGR \cite{cao2016deep}, and SDNE \cite{wang2016structural}. The algorithm in LINE consists of a combination of two encoder-decoder structures to study and optimize the first and second node proximities in the vector space. 
Both of the DNGR \cite{cao2016deep} and SDNE \cite{wang2016structural} algorithms embed the node local neighborhood information while using a random surfing method and approach single embeddings through auto-encoders than pairwise transformations. 

Although the graph representation learning models that are based on GNNs consider both graph structures and node features to embed the graph, they suffer from computational complexity and inefficiency in iterative updating of the hidden states. Furthermore, GNNs use the same parameters for all layers, which limits their flexibility. These architectures are always designed as shallow networks with no more than three layers, and including a higher number of layers is still being considered to be a challenge for CNNs \cite{zhou2018graph}.

The introduction of neural networks, specially convolutional neural networks, in order to graph structures, has led to extract features from complex graphs flexibly. Graph Convolutional Networks (GCNs) \cite{kipf2016semi} tackle the problem of high computational complexity and shallow architectures via defining a convolution operator for the graph. Furthermore, a rich class of convolutional filter functions can be achieved through stacking many convolution layers. The iterative aggregation of a node's local neighborhood is being used in GCNs to obtain graph embeddings, where this aggregation method leads to higher scalability besides learning graph global neighborhoods. The features for these models include the information from the topology of the network aggregated by the node attributes, when the node features are available from the data domain \cite{zhou2018graph}. Additionally, GCNs can be utilized for node embeddings, as well as subgraph embeddings \cite{hamilton2017representation}. Varying convolutional models have been derived from GCNs that employ different convolutional filters in their architecture. These filters can be designed as either spatial filters or spectral filters. The former type of convolutional filters can be directly operated on the original graph and its adjacency matrix; however, the latter type is being utilized on the spectrum of the graph Laplacian \cite{goyal2018graph}

In \cite{harada2018dual}, the problem of link prediction is studied while using a combination of two convolutional neural networks for the graph network of molecules. The molecules are represented as having a hierarchical structure for their internal and external interactions. The graph structure transformation to a low dimensional vector space is obtained from an internal convolutional layer that is randomly initialized for each node representation and trained by backpropagation. The external convolutional layer receives the embedded nodes as input to learn over the external graph representations. Finally,~the~link prediction algorithm consists of a multilayer neural network, which was accepting the final representations in order to predict the molecule-molecule interactions by a softmax function.

The algorithms that belong to the family of neighborhood aggregation methods, are also being referred to as convolutional models. An example is GraphSAGE \cite{hamilton2017inductive}, which aggregates the information from local neighborhoods recursively, or iteratively. This iterative characteristic leads the model to be generalizable to unseen nodes. The node attributes for this model might include simple node statistics, such as node degrees, or even textual data for profile information on online social networks. 

Graph convolutional neural networks for relational data analysis is proposed in \cite{schlichtkrull2018modeling}, which introduces Relational Graph Convolutional Networks (R-GCNs) for the task of link prediction and node classification. Because of relational models referring to directed associations, the node relationships in this models for the graph $G = \langle V, E, R \rangle$ are represented as $(v_x, r, v_y) \in E$, where $r \in R$ is a relation type for both canonical and inverse directions. This model can be considered to be a special case of simple differentiable message-passing model. In this model, the forward-pass update for entity $v_x$ in a relational multigraph can be propagated by:
\begin{equation}
    h_{v_x}^{(l+1)} = \sigma \bigg( \sum_{r \in R} \sum_{v_y \in \Gamma_{v_x}^r} \frac{1}{c_{v_x,r}} W_r^{(l)} h_{v_y}^{(l)} + W_0^{(l)} h_{v_x}^{(l)} \bigg),
\end{equation}
where $\sigma(.)$ is an element-wise activation function, $l$ denotes the layer of the neural network, $h_{v_x}$ is the hidden state of node $v_x$, $c_{v_x, r}$ refers to a problem-specific normalization constant, $W$ is the weight matrix, and $\Gamma_{v_x}^r$ denotes the set of neighbor indices of vertex $v_x$ under relation $r \in R$. Thus, this model is different from normal GCNs as the accumulation of transformed feature vectors of neighboring nodes are relation-specific. For this model, using multi-layer neural networks instead of simple linear message transformation is also possible. The task of link prediction by this model can be viewed as computing node representations with an R-GCN encoder and DistMult factorization \cite{yang2014embedding} as the scoring function, which is a known score function for relation representation with low number of relation parameters. The triple $(s, r, o)$ for (subject, relation, object) is being calculated in order to determine the likelihood of possible edges as:
\begin{equation}
    f(s,r,o) = v_{x_s}'^T R_r v_{x_o},
\end{equation}
in which $R_r \in \mathbb{R}^{d \times d}$ is a diagonal matrix for every relation $r$. The model can be trained with negative sampling via randomly corrupting the subject or object of positive examples. 

\section{Network Data Sets}
One of the challenging tasks in network research is the implementation and validation of the proposed methods and models. In the majority of the network research, the popular collections of data sets are used as common sense: a friendship network of 34 members of a Karate Club and 78 interactions among them \cite{zachary1977information}, the power network of an electrical grid of western US with 4941 nodes and 6594 edges \cite{watts1998collective}, an internet-based router network with 5022 nodes and 6258 edges \cite{spring2002measuring}, a protein--protein interaction network that contains 2617 proteins and 11855 interactions \cite{von2002comparative}, a collaboration network of 1589 authors with 2742 interactions \cite{newman2006finding}, an airline network of 332 nodes and 2126 edges that show the connection between airports (\url{http://vlado.fmf.uni-lj.si/pub/networks/data/}), a social network of 62 dolphins in New Zealand with 159 interactions \cite{lusseau2003bottlenose}, a biological network of the cerebral cortex of Rhesus macaque with 91 nodes and 1401 edges \cite{markov2014weighted}.

Data set collection is time-consuming and labor-intensive work. While some studies build their own data set, the researchers mostly prefer to employ an existing data set. Some popular collections of network data sets that might be used in link prediction studies are as follows:

\begin{itemize}[leftmargin=*,labelsep=3mm]
    \item SNAP \cite{snapnets}: a collection of more than 90 network data sets by Stanford Network Analysis Platform. With biggest data set consisting of 96 million nodes. 
    \item BioSNAP \cite{biosnapnets}: more than 30 Bio networks data sets by Stanford Network Analysis Platform
    \item KONECT \cite{Kunegis:2013:KKN:2487788.2488173}: this collection contains more than 250 network data sets of various types, including social networks, authorship networks, interaction networks, etc.
    \item PAJEK \cite{pajekdataset}: this collection contains more than 40 data sets of various types.
    \item Network Repository \cite{nr-aaai15}: a huge collection of more than 5000 network data sets of various types, including social networks.
    \item Uri ALON \cite{urialon}: a collection of complex networks data sets by Uri Alon Lab.
    \item NetWiki \cite{netwiki}: more than 30 network data sets collection of various types.
    \item WOSN 2009 Data Sets \cite{viswanath-2009-activity}: a collection of Facebook data provided by social computing group.
    \item Citation Network Data set \cite{Tang:08KDD}:  a collection of citation network dat aset extracted from DBLP, ACM, and other sources.
    \item Grouplens Research \cite{grouplens}: a movie rating network data set.
    \item ASU social computing data repository \cite{Zafarani+Liu:2009}: a collection of 19 network data sets of various types: cheminformatics, economic networks, etc.
    \item Nexus network repository \cite{nexus}: a repository collection of network data sets by iGraph.
    \item SocioPatterns \cite{sociopattern}: a collection of 10 network data sets that were collected by SocioPatterns interdisciplinary research collaboration.
    \item Mark Newman \cite{marknewman}: a collection of Network data sets by Mark Newman.
    \item Graphviz \cite{nr-aaai15}: an interactive visual graph mining and analysis.
\end{itemize}

\section{Taxonomy}
\begin{figure}[H]
\centering
		\includegraphics[width=1\linewidth, keepaspectratio]{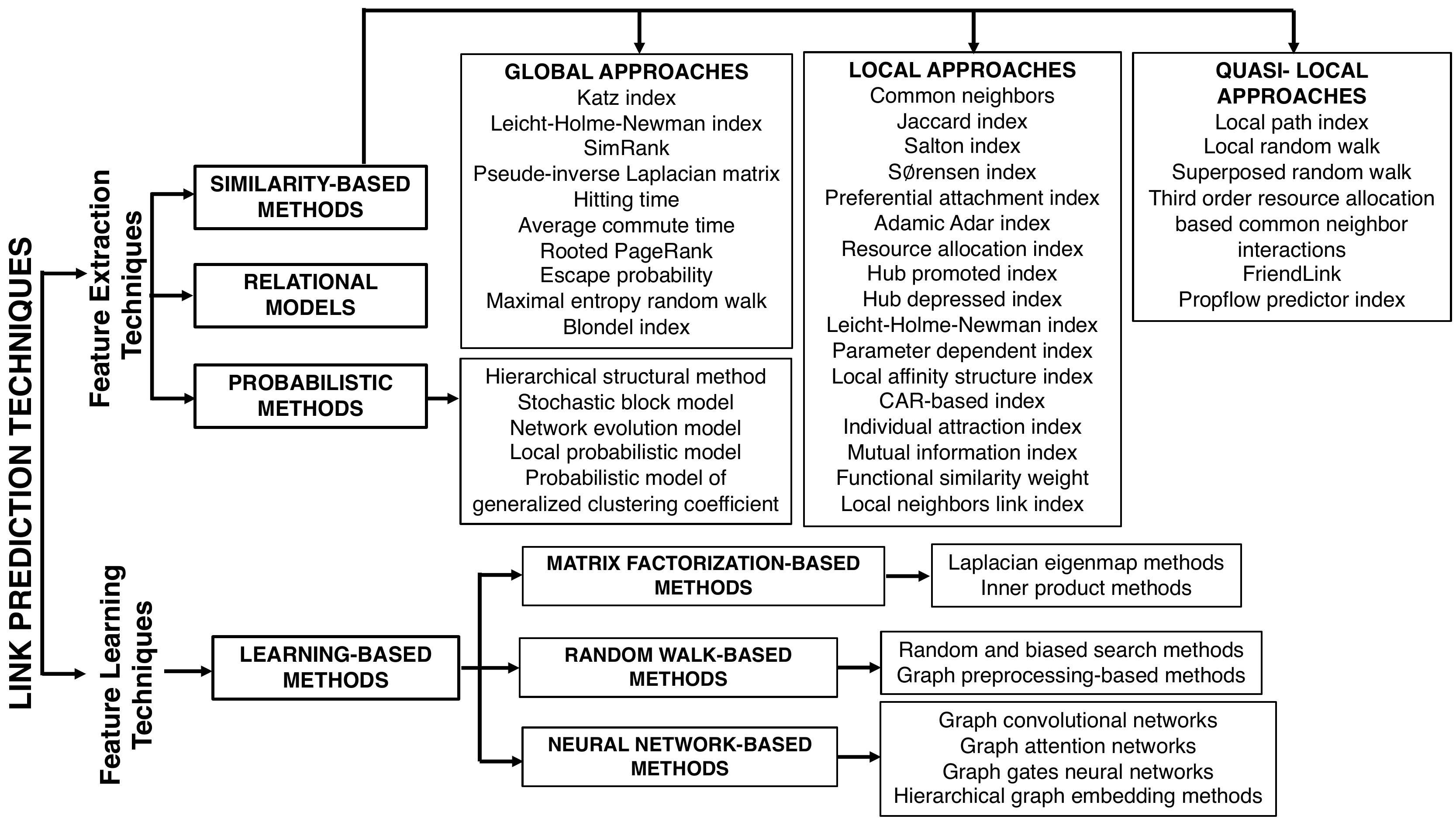}
		\caption{A taxonomy for the feature extraction techniques and feature learning methods in link prediction literature.
		}
		\label{fig:taxonomy}
\end{figure}

According to the methods that were explained earlier in this paper, we propose a taxonomy to better categorize the link prediction models. In our proposed taxonomy, the link prediction techniques are mainly categorized under two sections: feature learning and feature extraction techniques (Figure~\ref{fig:taxonomy}). 

\section{Discussion}
\label{sec:our_models}

This paper presents a comprehensive and state-of-the-art literature review on link prediction analysis, in which the emerging links or missed associations are predicted in a complex network, through a custom taxonomy. We classified link prediction techniques under two main categories. Firstly, feature extraction techniques consist of the methods that start with an initial set of features and build the required resources by using these raw features in order to describe the structural similarity. We discussed these methods under three different titles due to their strategy for addressing link prediction problems; namely, similarity-based, relational, and probabilistic methods. Among these methods, similarity-based techniques are the simplest and relatively less computationally intensive. These methods aim to explore missing links by assigning similarity scores between node pairs while using the structural properties of the graphs. According to the required topological information from a network, these methods are further divided into three subcategories. Global approaches require the complete topological information of the graph; therefore, they provide relatively more accurate results. However, the whole network may not be observable, or the large size of the network may require less time-consuming methods. In such cases, local approaches, in which a maximum second or third-degree neighborhood relationship is taken into consideration, rather than a whole network, are suggested to be applied instead. This trade-off triggered the emergence of the so-called quasi-local approaches. These methods are generally more favored and applied among similarity-based methods, since they are as efficient as global approaches due to the use of additional topological information, but less time-consuming. Other feature extraction techniques used in link prediction problems covered in this study are relational and probabilistic methods. Using maximum likelihood calculations in probabilistic methods makes them relatively time-consuming and expensive to deploy. Another major drawback of these models is the lack of the concept of an object, which is addressed in relational models. Thus, these models are able to use the logical structure of underlying data that is helpful for more complex problems. Accordingly, employing relational methods in a link prediction problem requires a massive computation of marginal probability distributions for each node in the network. Although these methods are considered to be powerful, the nonexistence of the compact closed form of these distributions due to mutual dependencies in the correlated networks makes their utilization challenging \cite{al2011survey}. Secondly, feature learning-based techniques consist of methods that allow for a system to automatically learn the necessary set of features before building the required resources to further address the link prediction problems. These high-performance approaches enable the integration of extra information that is related to the network that might be effective in predicting the existence of links, such as community structure \cite{mohan2017scalable}, users' behavior \cite{xiao20183}, common interests \cite{al2006link}, etc. Additionally, machine learning models are useful in picking the right combination of features by optimizing an objective function, which renders these methods more preferable when compared to the previously discussed approaches in many cases.   

Getoor and Diehl \cite{getoor2005link} categorized link prediction problems under four main sections: (i) the link existence, in which the likelihood of forming a connection between two nodes in the future is questioned, (ii) the link load, in which the weight of the associated links are analyzed, (iii) the link cardinality, in which the question of whether more than one link between a node pair exists or not is inspected, and (iv) the link type, in which the role of the link between node pairs are evaluated. Although the methods that are discussed in this survey mainly address the link prediction problem in networks, they can be easily prolonged to the problems of link load and link cardinality, since they both require a similar computational approach \cite{kushwah2016review}. Some learning-based methods and probabilistic models are being deployed for link prediction in temporal and dynamic networks. Whereas, the problem of link type differ since the prediction methods foTr multi-object type links may require special attention and the deployment of different methods. To obtain more detailed information regarding the commonly used approaches for link prediction problem in weighted, bipartite, temporal, dynamic, signed, and heterogeneous networks, please visit \cite{wind2012link,kunegis2010link,bu2019link,marjan2018link,daud2020applications,zhang2014link}, respectively.

Although the link prediction problem is an established field of research, several problems are yet to be explored in this domain. In general, the available methods in the literature produce new methods or compare the extant ones by assuming that the network is noise-free; however, some links might be missing, substituted, or fake, which is called noisy networks. While, Zhang et al. \cite{zhang2016measuring} compared a few numbers of similarity-based methods, but there is no detailed study that compare the robustness of different approaches. Besides, each network has its own characteristics, i.e., domain/network problem, and this makes transferring knowledge or generalizing the superiority of the link prediction algorithms challenging. Still, there are a few works that consider the effects of varying topological properties on the performance of different link prediction approaches. Furthermore, most of the real-world networks are shown to be sparse. The resulting unbalanced dataset obstructs the handling of link prediction problems, especially with the utilization of supervised techniques. Lastly, limited studies address the link prediction problem in multiplex/multilayer methods, and these studies are generally constrained with two layers. Further studies may consider this problem on multiplex networks with more than two layers.

\section*{Funding}
This work was supported by the Defense Advanced Research Projects Agency (DARPA) under grant number FA8650-18-C-7823.

\bibliographystyle{unsrt}  


\end{document}